\documentclass[
 reprint,
superscriptaddress,
nofootinbib,
 amsmath,amssymb,
 aps,
]{revtex4-2}

\usepackage{graphicx}
\usepackage{dcolumn}
\usepackage{bm}
\usepackage{hyperref}
\hypersetup{
    colorlinks=true,
    allcolors=blue
}
\usepackage{xcolor}
\usepackage{dsfont}

\usepackage{tabu}

\begin{document}

\title{Sub-Harmonic Control of a Fluxonium Qubit via a Purcell-Protected Flux Line}

\author{J.~Schirk}
\altaffiliation{These authors contributed equally to this work.}
\affiliation{Technical University of Munich, TUM School of Natural Sciences, Department of Physics, 85748 Garching, Germany}
\affiliation{Walther-Meißner-Institut, Bayerische Akademie der Wissenschaften, 85748 Garching, Germany}
\author{F.~Wallner}
\altaffiliation{These authors contributed equally to this work.}
\affiliation{Technical University of Munich, TUM School of Natural Sciences, Department of Physics, 85748 Garching, Germany}
\affiliation{Walther-Meißner-Institut, Bayerische Akademie der Wissenschaften, 85748 Garching, Germany}
\author{L.~Huang}
\author{I.~Tsitsilin}
\author{N.~Bruckmoser}
\author{L.~Koch}
\author{D.~Bunch}
\author{N. J.~Glaser}
\author{G. B. P.~Huber}
\author{M.~Knudsen}
\author{G.~Krylov}
\affiliation{Technical University of Munich, TUM School of Natural Sciences, Department of Physics, 85748 Garching, Germany}
\affiliation{Walther-Meißner-Institut, Bayerische Akademie der Wissenschaften, 85748 Garching, Germany}
\author{A.~Marx}
\affiliation{Walther-Meißner-Institut, Bayerische Akademie der Wissenschaften, 85748 Garching, Germany}

\author{F.~Pfeiffer}
\author{L.~Richard}
\affiliation{Technical University of Munich, TUM School of Natural Sciences, Department of Physics, 85748 Garching, Germany}
\affiliation{Walther-Meißner-Institut, Bayerische Akademie der Wissenschaften, 85748 Garching, Germany}
\author{F. A.~Roy}
\affiliation{Walther-Meißner-Institut, Bayerische Akademie der Wissenschaften, 85748 Garching, Germany}
\affiliation{Theoretical Physics, Saarland University, 66123 Saarbrücken, Germany}
\author{J. H.~Romeiro}
\author{M.~Singh}
\author{L.~Södergren}
\affiliation{Technical University of Munich, TUM School of Natural Sciences, Department of Physics, 85748 Garching, Germany}
\affiliation{Walther-Meißner-Institut, Bayerische Akademie der Wissenschaften, 85748 Garching, Germany}
\author{E.~Dionis}
\author{D.~Sugny}
\affiliation{Laboratoire Interdisciplinaire Carnot de Bourgogne (ICB), UMR 6303 CNRS-Université de Bourgogne, 9 Av. A. Savary, BP 47 870, F-21078 Dijon, France}
\author{M.~Werninghaus}
\author{K.~Liegener}
\author{C. M. F.~Schneider}
\email{christian.schneider@wmi.badw.de}
\affiliation{Technical University of Munich, TUM School of Natural Sciences, Department of Physics, 85748 Garching, Germany}
\affiliation{Walther-Meißner-Institut, Bayerische Akademie der Wissenschaften, 85748 Garching, Germany}
\author{S.~Filipp}
\affiliation{Technical University of Munich, TUM School of Natural Sciences, Department of Physics, 85748 Garching, Germany}
\affiliation{Walther-Meißner-Institut, Bayerische Akademie der Wissenschaften, 85748 Garching, Germany}
\affiliation{Munich Center for Quantum Science and Technology (MCQST), 80799 München, Germany}
\date{\today}

\begin{abstract}
Protecting qubits from environmental noise while maintaining strong coupling for fast high-fidelity control is a central challenge for quantum information processing. Here, we demonstrate a control scheme for superconducting fluxonium qubits that eliminates qubit decay through the control channel by suppressing the environmental density of states at the transition frequency. Adding a low-pass filter on the flux line allows for flux-biasing and, at the same time, coherently controlling the fluxonium qubit by parametrically driving it at integer fractions of its transition frequency. We compare the filtered to the unfiltered configuration and find a five times longer $T_1$, and ten times improved $T_2$-echo time in the filtered case.
We demonstrate coherent control with up to 11-photon sub-harmonic drives, highlighting the strong non-linearity of the fluxonium potential.  
Measured Rabi frequencies and drive-induced frequency shifts show excellent agreement with numerical and analytical models.
Furthermore, we show the equivalence of a 3-photon sub-harmonic drive to an on-resonance drive by benchmarking sub-harmonic gate fidelities above 99.94$\,$\%. 
These results open up a scalable path for full qubit control through a single Purcell-protected channel, providing strong suppression of control-induced decoherence and enabling wiring-efficient superconducting quantum processors.
\end{abstract}

\maketitle

\section{\label{introduction}Introduction}
Superconducting circuits are a promising platform for scalable, error-corrected quantum processors~\cite{krinnerRealizingRepeatedQuantum2022,googlequantumaiSuppressingQuantumErrors2023,sivakRealtimeQuantumError2023}. 
This potential has been enabled by collective efforts in improving circuits based on transmon qubits~\cite{kochChargeinsensitiveQubitDesign2007}, culminating in demonstrating quantum algorithms close to practical utility~\cite{kimEvidenceUtilityQuantum2023}. 
One of the main limitations for current superconducting quantum processors is decoherence~\cite{googlequantumaiSuppressingQuantumErrors2023}, which sets an upper bound on achievable gate fidelities.
To improve further, it is crucial to understand, characterize, and close all channels contributing to the loss of quantum information~\cite{lisenfeldElectricFieldSpectroscopy2019, gyenisMovingTransmonNoiseProtected2021}. 

Decoherence channels can be classified into two categories: internal and external losses~\cite{gaoPhysicsSuperconductingMicrowave2008}.
Internal losses arise from material imperfections such as two-level systems (TLS) that couple to the qubit~\cite{lisenfeldElectricFieldSpectroscopy2019, degraafDirectIdentificationDilute2017, spieckerTwolevelSystemHyperpolarization2023a}. 
By optimizing designs and fabrication processes, it is possible to significantly reduce both the coupling to TLSs and their density~\cite{wangSurfaceParticipationDielectric2015, crowleyDisentanglingLossesTantalum2023}. 
In this regard, the fluxonium qubit~\cite{somoroffMillisecondCoherenceSuperconducting2023,manucharyanFluxoniumSingleCooperPair2009} stands out demonstrating record-high coherence times~\cite{somoroffMillisecondCoherenceSuperconducting2023,wangHighcoherenceFluxoniumQubits2025,dingHighFidelityFrequencyFlexibleTwoQubit2023} among superconducting qubits attributed to its weaker coupling to TLS at low transition frequencies of a few hundred MHz~\cite{linDemonstrationProtectionSuperconducting2018a}. 
Recently, the single-qubit and two-qubit gate performance of fluxonium qubits~\cite{moskalenkoHighFidelityTwoqubit2022, dingHighFidelityFrequencyFlexibleTwoQubit2023, zhangTunableInductiveCoupler2024} has surpassed the more widely used transmon qubits, thus providing a viable alternative for superconducting quantum processors~\cite{nguyenBlueprintHighPerformanceFluxonium2022}. 

External sources of decoherence result from signal lines that are required to control the qubits~\cite{houckControllingSpontaneousEmission2008, schoelkopfQubitsSpectrometersQuantum2003, krinnerEngineeringCryogenicSetups2019, yanFluxQubitRevisited2016}.
One approach to mitigate their impact is to reduce the coupling between the qubit and the control lines, while compensating with increased drive power. However, this strategy is ultimately constrained by the limited cooling power available in state-of-the-art cryogenic systems~\cite{krinnerEngineeringCryogenicSetups2019}.
As a result, fast and high-fidelity control necessitates a finite, sufficiently strong coupling to a control line~\cite{konoBreakingTradeoffFast2020}. 
However, since linear couplings are reciprocal, the control line inevitably acts as a decay channel for the qubit. 
Ideally, one would like to independently engineer the coupling for qubit control from the qubit dissipation into the control channel~\cite{konoBreakingTradeoffFast2020}. 
Breaking the reciprocity of these processes requires exploiting non-linear effects like multi-photon processes, which are observed in non-linear media where driven systems interact with higher harmonics~\cite{shirleySolutionSchrodingerEquation1965, ferrayMultipleharmonicConversion10641988,mcphersonStudiesMultiphotonProduction1987}, or with sub-harmonics at integer fractions of the transition frequency~\cite{nakamuraRabiOscillationsJosephsonJunction2001, wallraffMultiphotonTransitionsEnergy2003, saitoMultiphotonTransitionsMacroscopic2004, liuSingleFluxQuantumBased2023,xiaFastSuperconductingQubit2023,sahDecayprotectedSuperconductingQubit2024}. 
Fluxonium qubits, with their strong non-linear potential, are well suited for multi-photon control processes. 
Interestingly, such multi-photon processes can be driven directly via an inductively coupled flux bias line. This same line serves to maintain the fluxonium at its ideal operation point, where it is first order insensitive to flux noise. 
In addition to minimizing the number of control elements, resonant flux control has been shown to yield higher single-qubit gate fidelities than charge-based control in fluxonium qubits, making it especially attractive for implementing multi-photon gates~\cite{rowerSuppressingCounterRotatingErrors2024}. To suppress noise at the qubit transition frequency and reduce decoherence introduced by the control line, a low-pass filter can be inserted into the flux line.
In this configuration, sub-harmonic control enables full qubit control over a single filtered channel, eliminating the need to resonantly drive the qubit at its transition frequency over a separate charge line. 

In this work, we demonstrate sub-harmonic control of a fluxonium qubit using a control channel with a low-pass filter below the fundamental qubit frequency. 
We confirm that the low-pass filter isolates the qubit from external noise when idling, resulting in a fluxonium primarily limited by internal losses and readout resonator population. 
Subsequently, we demonstrate coherent control of the fluxonium qubit through the filtered channel using sub-harmonic driving. 
This results in single qubit gates with fidelities above $\mathcal{F}>99.9$\,\%.

\section{\label{protection} Purcell Protection}

\begin{figure}
    \centering
    \includegraphics[width=0.99\linewidth]{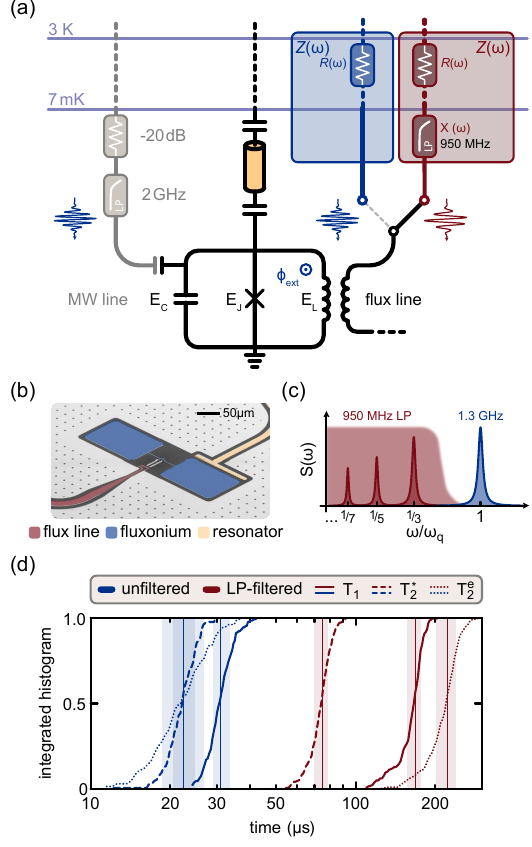}
    \caption{\textbf{Setup and relaxation times.} (a) Setup at the low temperature stages of the cryostat (1\,K and 100\,mK stages are left out and displayed in Appedix~\ref{fig:setup_full}) , including readout resonator and two switchable flux line configurations (unfiltered (UF) and low-pass filtered (LP)). An additional microwave (MW) line is added for comparison and benchmarking.
    (b) False-color micrograph of the fluxonium (blue) with a galvanically coupled flux line (red) and a capacitively coupled readout resonator (yellow). The additional MW line is located outside of the micrograph cut-out shown here.  (c) Sketch of the power spectral density  $S(\omega)$ for the resonant (blue) and sub-harmonic drive (red) setup. The shaded area indicates the pass band of the low-pass filter.
    (d) Integrated histograms for $T_1$ (solid), $T_2^*$ (dashed), and $T_2^\mathrm e$ (dotted) for unfiltered (blue) and filtered (red) flux line configuration. Vertical lines with shaded areas illustrate the median and standard deviation with values given in Table~\ref{coherence_table}.
    }
    \label{fig:fig1}
\end{figure}

Qubit dissipation into its environment is described by Fermi's golden rule.
For an environment with power spectral density $S(\omega)$ coupling to the qubit via the transition operator $\hat A$, the relaxation rate $\gamma_\mathrm{eg}$ between the qubit ground state $|g\rangle$ and first excited state $|e\rangle$ is given by~\cite{schoelkopfQubitsSpectrometersQuantum2003, kochChargeinsensitiveQubitDesign2007, popCoherentSuppressionElectromagnetic2014a}
\begin{equation}
    \gamma_\mathrm{eg} = \frac{1}{\hbar^2}|\langle g|\hat A|e\rangle|^2 S(\omega_\mathrm{eg}),
    \label{eq:fermiGolden}
\end{equation}
where $\hbar\omega_\mathrm{eg}$ is the energy separation of the qubit states.
The power spectral density of an environment inductively coupled to a superconducting circuit is given by the fluctuation-dissipation theorem~\cite{nyquistThermalAgitationElectric1928, popCoherentSuppressionElectromagnetic2014a}   
\begin{equation}
    S(\omega)=\hbar\omega\frac{R(\omega)}{|Z(\omega)|^2}\left[1+\coth\left(\frac{\hbar\omega}{2k_\mathrm{B}T}\right)\right].
    \label{eq:noisedensity}
\end{equation}
Here the first term describes spontaneous emission and the second term induced emission and absorption processes. 
We model the environment without loss of generality as a parallel current noise source with impedance $Z(\omega)$. 
Similarly, the environment can be modeled as a parallel noise voltage source with capacitive coupling~\cite{houckControllingSpontaneousEmission2008, reedEntanglementQuantumError2013}.
The impedance consists of a resistive component $R(\omega)=\mathrm{Re}[Z(\omega)]$ at temperature $T$, and reactive components $X(\omega)=\mathrm{Im}[Z(\omega)]$. The latter can be composed of inductive and capacitive elements leading to a frequency-dependent reactance and, therefore, a filter transfer function. 

According to Eq.~(\ref{eq:fermiGolden}), qubit dissipation is suppressed when the matrix element $\langle g|\hat A|e\rangle$ vanishes, as is the case in protected circuits such as the zero-pi~\cite{gyenisExperimentalRealizationProtected2021} or inductively-shunted transmon~\cite{hassaniInductivelyShuntedTransmons2023}. 
In circuits with non-vanishing matrix elements, qubit dissipation can instead be reduced by lowering the noise spectral density $S(\omega_\mathrm{eg})$ according to Eq.~(\ref{eq:noisedensity}).
This reduction can be achieved by lowering the temperature, since the $\coth$ term approaches unity for $T\to0$, minimizing stimulated emission.
This is typically achieved in qubits with transition frequencies above 5\,GHz by state-of-the-art dilution cryostats at millikelvin temperatures~\cite{krinnerEngineeringCryogenicSetups2019}. For fluxonium qubits with transition frequencies below 1\,GHz, this approximation is no longer valid due to the exponential increase of the $\coth$ term for $\hbar\omega/2k_\mathrm BT<1$. 

However, the inclusion of reactive elements offers another possibility to suppress environment-induced qubit relaxation by modifying the complex impedance $Z(\omega)$ at the qubit frequency $\omega_\mathrm{eg}$. 
This approach affects both the stimulated and spontaneous emission term in Eq.~(\ref{eq:noisedensity}) and is referred to as Purcell-protection~\cite{purcellResonanceAbsorptionNuclear1946}. 
We can achieve a vanishing power spectral density at the qubit frequency $S(\omega_\mathrm{eg})\to0$ by either $R(\omega_\mathrm{eg})\to0$ for finite $|Z(\omega_\mathrm{eg})|$ or $|Z(\omega_\mathrm{eg})|\to\infty$ for finite $R(\omega_\mathrm{eg})$. The former describes a low-pass filter above the cut-off frequency and the latter a high-pass filter below the cut-off frequency. As fluxonium qubits require a DC bias for its ideal operation point, we will focus on low-pass filters in the following. Intuitively, the reduction of the decay rate for low-pass filters occurs because a vanishing resistance results in smaller fluctuations according to the fluctuation-dissipation theorem.

While a low-pass filter below the qubit frequency $\omega_\mathrm{eg}$ results in a vanishing contribution of the control-line to $\gamma_\mathrm{eg}$, it prevents direct driving of the qubit at $\omega_\mathrm{d} =\omega_\mathrm{eg}$. However, through sub-harmonic control, as described below in Section \ref{parametic_chapter}, we decrease the drive frequency $\omega_\mathrm{d}$ to a regime well below the cut-off frequency of the filter, restoring qubit control capabilities while retaining low qubit dissipation.

 We first demonstrate the suppression of qubit dissipation through filtering at the qubit frequency on a fluxonium qubit consisting of a single junction with Josephson energy $E_\mathrm{J}$, shunted by capacitive pads with a charging energy $E_\mathrm{C}$ and an inductance with inductive energy $E_\mathrm{L}$, as shown in Fig.~\ref{fig:fig1}\,(a). The fluxonium circuit is described by the Hamiltonian~\cite{youCircuitQuantizationPresence2019}
\begin{equation}
    \hat{H} = 4E_\mathrm{C}\hat{n}^2 - E_\mathrm{J}\cos{\hat{\varphi}} + \frac{E_\mathrm{L}}{2}\left[\hat{\varphi} - \phi(t)\right]^2,
    \label{eq:hamiltonian}
\end{equation}
with the reduced phase and charge operators $\hat{\varphi}$, $\hat{n}$ and the external flux $\phi(t)=2\pi\Phi(t)/\Phi_0$ in units of the magnetic flux quantum $\Phi_0$.  
A false-color micrograph of the device is shown in Fig.~\ref{fig:fig1}\,(b), and further details of the circuit and its parameters are listed in Appendix~\ref{appendix:qubit-params}. 
We include both an inductively-coupled flux line to control the external flux and the qubit states as well as an auxiliary weakly-coupled microwave (MW) line as a reference for controlling the qubit states.
The flux line couples to the fluxonium via the phase operator $\hat{\varphi}$ and the MW-line via the charge operator $\hat{n}$. Since the MW-line is only weakly coupled and strongly attenuated, we neglect it in the following discussion. 
 
In order to modify the impedance of the flux line, we insert a low-pass filter (\textit{Minicircuits} VLFX780+) with a 3\,dB cut-off frequency at 950\,MHz, which is well below the qubit frequency $\omega_\mathrm{eg}/2\pi=1.32\,$GHz [Fig.~\ref{fig:fig1}\,(c)]. 
The filter decreases $R(\omega_\mathrm{eg})/|Z(\omega_\mathrm{eg})|^2$ to reduce relaxation according to Eq.~(\ref{eq:noisedensity}). Below the cut-off frequency, it maintains at $R(\omega_\mathrm{eg})/|Z(\omega_\mathrm{eg})|^2\approx1/50\,\Omega$ to enable fast control operations.
To assess the effect of the filtered environment on qubit performance, we excite the qubit resonantly over the MW line and measure the relaxation ($T_1$), Ramsey coherence ($T_2^*$) and Hahn-echo times ($T_2^{\mathrm{e}}$) of the fluxonium qubit separately for the unfiltered (UF) and low-pass filtered  (LP) flux line configuration.  
Our setup includes a cryogenic microwave switch [Fig. \ref{fig:fig1}\,(a)], to change in-situ between both configurations.
We measure the coherence times in both configurations for several hours. The results are plotted as integrated histograms in Fig. \ref{fig:fig1}\,(d) with the median values depicted in Table \ref{coherence_table}.
Adding the filter significantly increases average relaxation times by a factor of five up to $T_1=168(20)$\,\textmu s and dephasing times by a factor three and ten up to $T_2^*=75(9)$\,\textmu s and $T_2^\mathrm{e}=223(37)$\,\textmu s, respectively. 

In addition, we extract the effective qubit temperature $T_{\mathrm{eff}}$ from single-shot measurements (see Appendix \ref{appendix:temperatures} for details).
We observe a decrease in the effective temperature from $T_\mathrm{eff}=245(25)$\,mK without the LP filter (UF) to $T_\mathrm{eff}=28.9(5)$\,mK with the filter (LP).
These results confirm that the relaxation time for the unfiltered line setup is dominantly limited by stimulated emission due to thermal photons, with $T_1 \propto 1/(2n_{\mathrm{th}}+1)$~\cite{blaisCircuitQuantumElectrodynamics2021}. 
For further verification, we extract the coupling of the qubit to the flux line represented as a bath at different temperatures.
With the measured power transmission factor of approximately $-35.5\,$dB for the low-pass filter at $\omega_\mathrm{eg}$ and assuming that the lowest temperature attenuator of the flux line is well thermalized to 3\,K, we determine a coupling rate of $\gamma_\mathrm{flux}=1/3.6(5)$\,ms for the flux line as detailed in Appendix \ref{appendix:temperatures}. This value agrees well with the design value of $\gamma_\mathrm{flux}=1/3.4$\,ms.

\begin{table}
\begin{center}
\begin{tabular}{c|c|c|c}  
 & \textbf{unfiltered}  & \textbf{LP-filtered} & \textbf{improvement} \\ 
 \hline
 $T_1$ & 31(5)\,\textmu s & 168(20)\,\textmu s & $\times$\,5\,(1)\\ 
 $T_2^*$ & 22(4)\,\textmu s & 75(9)\,\textmu s & $\times$\,3\,(1)\\ 
 $T_2^\mathrm e$ & 22(8)\,\textmu s & 223(37)\,\textmu s & $\times$\,10\,(4)\\
 $T_\mathrm{eff}$ & 245(25)\,mK & 28(1)\,mK & $\times$\,9\,(1) \\
\end{tabular}
\caption{\textbf{Decoherence times and effective qubit temperature for the filtered and unfiltered setup.}\label{coherence_table} 
Energy relaxation $T_1$, Ramsey-coherence $T_2^*$, Hahn-echo $T_2^\mathrm e$, and effective temperature $T_\mathrm{eff}$ are listed for the unfiltered and low-pass filtered flux line configuration. 
}
\end{center}
\end{table}
Furthermore, in the unfiltered configuration, we observe almost equal values for the Ramsey coherence $T_2^*$ and Hahn-echo times $T_2^\mathrm{e}$.
This suggests that high-frequency noise is the dominant source for decoherence in this configuration, since it cannot be compensated by the low-frequency noise insensitive Hahn-echo sequence. In the low-pass filtered case we observe a large difference between $T_2^\mathrm{e}$ and $T_2^*$, pointing towards low-frequency noise as the primary decoherence mechanism~\cite{yanRotatingframeRelaxationNoise2013}.
Given that the fluxonium qubit is first-order insensitive to flux noise at its sweet spot, we attribute the observed dephasing primarily to thermal photons in the readout resonator. 
By using a dispersive model for the qubit-resonator system described in Appendix~\ref{appendix:T2Limit}, we estimate an effective resonator temperature of 51(1)\,mK, which is in good agreement with temperatures reported in other works~\cite{ardatiUsingBifluxonTunneling2024,somoroffMillisecondCoherenceSuperconducting2023,yanDistinguishingCoherentThermal2018, pfeifferEfficientDecouplingNonlinear2024}. This indicates that coherence is not limited by flux noise through the passband of the low-pass filter, but rather by the strong coupling between readout resonator and qubit. 
\section{Parametric sub-harmonic driving}\label{parametic_chapter}
While filtering is an effective method to protect the idling qubit from relaxation caused by control lines, it impedes control over the qubit state using drive pulses at its transition frequency $\omega_\mathrm{eg}$. 
This limitation is overcome by driving the qubit at an integer fraction $\omega_\mathrm{eg}/n$ of its frequency, i.e. a $n$-photon sub-harmonic drive, which excites the qubit.

\begin{figure*}
    \centering
    \includegraphics{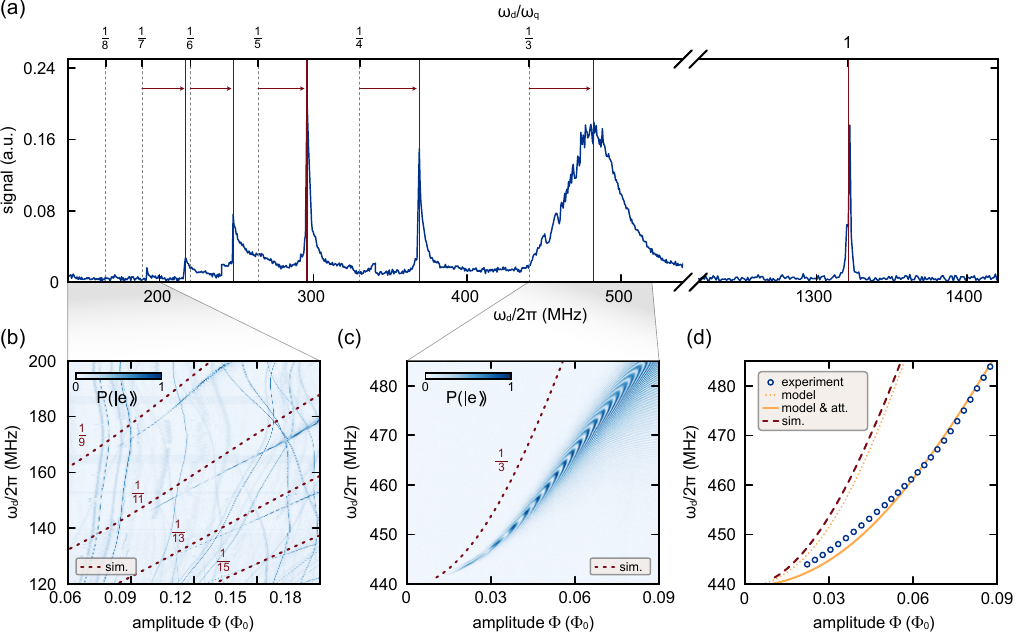}
    \caption{\textbf{Pulsed sub-harmonic qubit spectroscopy.} (a) Qubit spectroscopy at a fixed amplitude. We observe qubit excitation at its transition frequency $\omega_\mathrm{eg}$ and at fractions $1/n$ of $\omega_\mathrm{eg}$. The frequency range around $\omega_\mathrm{eg}$ is measured by driving over the auxiliary MW-line at -30\,dB reduced power.
    (b), (c) Measured excited state population of the qubit under sub-harmonic driving for varying drive amplitude and frequency ranges around $\omega_\mathrm{eg}/3$ (c) and below $\omega_\mathrm{eg}/9$ (b). Dashed lines indicate the simulated resonant drive-frequency of the sub-harmonic transitions.
    (d) Resonant drive frequency of the $3^{\rm rd}$ sub-harmonic as a function of drive amplitude for data (blue circles), model (orange dashed line) and simulations (red dashed line). Including a frequency-dependent attenuation of the flux bias line results in a good agreement between model and data (orange solid line).}
    \label{fig:fig2}
\end{figure*}

To investigate this process for fluxonium qubits, we bias the qubit at its lower sweet spot and apply a time-dependent drive to the flux line. 
The corresponding Hamiltonian in Eq.~(\ref{eq:hamiltonian}) can be separated into a time-independent part $\hat H_0$ and a time-dependent drive term
$\hat H_\mathrm{d}$
, given by
\begin{align}
    &\hat H_0= 4E_\mathrm{C}\hat{n}^2 + E_\mathrm{J}\cos{\hat{\varphi}} + \frac{E_\mathrm{L}}{2}\hat{\varphi}^2
    \label{eq:rest_hamiltonian}
    \\
    & \hat H_\mathrm{d}(t)=-E_\mathrm{L}\phi(t)\hat{\varphi},
    \label{eq:driven_hamiltonian}
\end{align}  
where we omitted global energy offsets (the sign change in Eq.~(\ref{eq:rest_hamiltonian}) results from operating the qubit at its half-integer flux sweet spot where $\cos(\hat\varphi+\pi)=-\cos\hat\varphi$).
The time dependent flux modulation has the form
\begin{equation}
    \phi(t) = 2\pi\frac{\Phi}{\Phi_0} E(t)\cos(\omega_\mathrm{d}t),
    \label{eq:pulse_shape}
\end{equation}
with the drive frequency $\omega_\mathrm{d}$, the pulse amplitude $\Phi$ and a normalized pulse envelope $E(t)$.
In experiment, we probe the excitation spectrum with a pulsed qubit spectroscopy by sweeping the drive frequency $\omega_d$ from  $\omega_\mathrm{eg}/8$  to $\omega_\mathrm{eg}/3$  and measuring the excited state population $P(|e\rangle)$ as shown in Fig.~\ref{fig:fig2}\,(a).
This frequency range is selected to probe the most dominant multi-photon transitions by using a flat-top Gaussian pulse with drive amplitude $\Phi/\Phi_0=0.083$ and a pulse length of 850\,ns. 
We observe multiple signal peaks at frequencies close to integer fractions of $\omega_\mathrm{eg}$.
However, all transitions show a systematic shift to higher frequencies, making it challenging to unambiguously identify signal peaks with their respective photon number.

To quantitatively assess the drive induced frequency shift, we repeat the spectroscopy with varying drive amplitude $\Phi$ around the frequency of the $3^\mathrm{rd}$ sub-harmonic, as well as for a lower frequency range where transitions with higher photon numbers are expected. 
The resulting spectroscopy for the $3^\mathrm{rd}$ sub-harmonic, displayed in Fig.~\ref{fig:fig2}\,(c), shows that the frequency shift increases monotonically with $\Phi$ and converges to $\omega_\mathrm{eg}/3$ at low drive powers. 
The low-frequency spectroscopy shown in Fig.~\ref{fig:fig2}\,(b) features similar transitions that increase monotonically with amplitude, as well as various transitions that exhibit a non-trivial dependence on amplitude. We attribute these transitions to the drive-induced coupling of the qubit to other systems, such as the readout-resonator, defects on the surface of the chip~\cite{mullerUnderstandingTwolevelsystemsAmorphous2019} or other parasitic resonances.
A detailed study of the origin of the transitions is, however, beyond the scope of this work. 
We identify the spectral lines corresponding to a multi-photon drive by comparing the spectrum to numerical simulations of the time evolution of the ground state $|g\rangle$ under the Hamiltonian $\hat H_0+\hat H_\mathrm d(t)$ (see Appendix \ref{appendix:simulation}). 
Remarkably, we find spectral lines for 9-, 11-, 13- and 15-photon transitions (red dashed lines in Fig.~\ref{fig:fig2}), which highlights the strong non-linearity of the fluxonium potential.
All measured spectral resonances show a distinct shift in frequency compared to the power scaling predicted in simulations. 

In order to better understand the observed behavior, we derive an effective analytical model for the $n$-photon sub-harmonic drive. We start with Eq.~(\ref{eq:rest_hamiltonian}), and express the undriven Hamiltonian $\hat H_0$ in terms of its eigenstates $|m\rangle$:
\begin{equation}
   \hat H_0 = \hbar\omega_\mathrm{eg}\hat b^\dagger \hat b + \frac{\hbar \alpha}{2} \hat b^\dagger \hat b^\dagger \hat b \hat b + \cdots,
    \label{eq:h0}
\end{equation}
where we introduced the ladder operator $\hat b$ defined as $\hat b|m\rangle=\sqrt{m}|m-1\rangle$ and the anharmonicity $\alpha$ of the qubit\footnote{We emphasize that $\hat b^\dagger$ and $\hat b$ are not equal to the ladder operators of an $LC$ oscillator, for which the flux operator can be expressed as $\hat\varphi\propto\hat a+\hat a^\dagger$. This approximation only holds for weakly anharmonic oscillators such as the transmon \cite{blaisCircuitQuantumElectrodynamics2021}.}. 
We emphasize that infinitely many terms in Eq.~(\ref{eq:h0}) are needed to adequately describe the fluxonium spectrum. Given that our fluxonium is less anharmonic than conventional implementations (c.f. Appendix~\ref{appendix:qubit-params}), we neglect all but the first two terms, resulting in a three-level approximation, which is equivalent to numerical simulations that include the lowest 20 states and is sufficient to explain the observed three-photon behavior quantitatively. 
Expressing the flux operator $\hat\varphi$ in terms of $\hat b^\dagger$ and $\hat b$, $\hat H_\mathrm{d}(t)$ takes the form
\begin{equation}
    \hat H_\mathrm{d} =-\phi(t)E_\mathrm{L}\hat\varphi = -\phi(t)[\beta_1(\hat b+\hat b^\dagger)+\beta_2(\hat b^\dagger \hat b \hat b+\hat b^\dagger \hat b^\dagger\hat b)],
    \label{eq:ht}
\end{equation}
with the coefficients $\beta_1= E_\mathrm{L} \langle g| \hat\varphi |e\rangle$ and, $\beta_2=\frac{E_\mathrm{L}}{\sqrt{2}} \langle e| \hat\varphi |f\rangle-\beta_1$, which determine the drive strength.

The emergence of $n$-photon transitions from $\hat H_0+\hat H_\mathrm{d}(t)$ is described perturbatively by a Magnus expansion~\cite{magnusExponentialSolutionDifferential1954} of the time evolution
\begin{equation}
    \hat U(0,t_\mathrm{pulse}) = \mathcal{T}\exp\left(-i \int_0^{t_\mathrm{pulse}} dt \hat H(t) \right),
\end{equation}
where $\mathcal{T}$ denotes time-ordering operator.
Each expansion order is proportional to time integrals over the times $t_i$ of the nested commutators $[\hat H(t_1), [\hat H(t_2),[\cdots]]]$. A multi-photon drive arises only if $\hat H(t)$ contains terms where the nested commutator is proportional to $\hat b$ or $\hat b^\dagger$ and its oscillation period is on-resonance with the shifted transition $\omega_\mathrm{eg}+\delta$ of the driven fluxonium.
When using bosonic operators, one can easily identify two operator pairs that satisfy these conditions for a drive frequency $\omega_\mathrm{d}\sim\omega_\mathrm{eg}/n$.
The first non-commuting pair is the $\alpha$-term in $\hat{H}_0$ and the linear $\beta_1$-term in $\hat H_\mathrm{d}$, i.e. when $n=3$, $\beta_1^3[\hat{b}^\dagger,[\hat{b}^\dagger,[\hat{b},\frac{\alpha}{2}\hat{b}^\dagger\hat{b}^\dagger\hat{b}\hat{b}]]]\propto\hat{b}^\dagger$. It can be understood as the drive term interacting with the anharmonicity of the fluxonium.
Furthermore, the second pair contains two drive terms proportional to $\beta_1$ and $\beta_2$ from $\hat H_\mathrm{d}$, i.e. when $n=3$, $\beta_1^2\beta_2[\hat{b}^\dagger,[\hat{b},\hat{b}^\dagger\hat{b}^\dagger\hat{b}]]]\propto\hat{b}^\dagger$. It can be interpreted as a self-interaction term of the drive operator. 
The resulting effective Hamiltonian for the $n$-photon drive, derived in detail in Appendix~\ref{appendix:theory}, is given by
\begin{align}
\begin{aligned}
    \frac{\hat{H}_{\mathrm{eff}, n}}{\hbar} = &(\omega_\mathrm{eg} +\delta_n - n\omega_\mathrm{d}) \hat b^\dagger \hat b + \frac{\alpha + \alpha_n}{2} \hat b^\dagger \hat b^\dagger \hat b \hat b \\
    &+ \Omega_n (\hat b^\dagger + \hat b).
\end{aligned}
\end{align}
The drive induces a frequency shift on all energy eigenstates, here visible as $\delta_n$ and $\alpha_n$ for the first and second excited states $|m=1\rangle$ and $|m=2\rangle$, respectively.
Additionally, $\hat{H}_{\mathrm{eff}, n}$ contains an effective coupling term $\Omega_n$ equal to the Rabi frequency when driven at the sub-harmonic frequency $\omega_\mathrm{d}=(\omega_\mathrm{eg}+\delta_n)/n$. 
The leading order contribution to the frequency shift, denoted as \(\delta_n^{(2)}\), arises from a second-order process related to self-interaction, and its proportionality is given by
\begin{equation}
\delta_n^{(2)} \propto \frac{n}{(n-1)^2\omega_\mathrm{d}}\left[\langle g|\hat\varphi|e\rangle^2-\frac{1}{2}\langle e|\hat\varphi|f\rangle^2\right]\left(\frac{\Phi}{\Phi_0}\right)^2.
\end{equation}
Thus, we can determine the sign of $\delta_n$ from the matrix elements of the drive operator $\hat\varphi$, and for our device we expect positive frequency shifts.
Note that for more anharmonic fluxonium qubits, $\delta_n$ increases strongly since $\omega_\mathrm{d}$ decreases and the matrix element asymmetry, here $\beta_2$, increases.
In comparison, the matrix asymmetry vanishes in transmon qubits, where $\langle e|\hat\varphi|f\rangle = \sqrt 2\langle g|\hat\varphi|e\rangle$ or equivalently $\beta_2=0$, thus only higher order contributions determine the frequency shifts.
In general, the Rabi-frequency for an n-photon process $\Omega_n$ scales with the drive amplitude $\Phi$ according to a power law $\Omega_n\propto\Phi^n$, as derived in Appendix~\ref{appendix:theory}.
Similar to the expression for the frequency shift \(\delta_n^{(2)}\) the dominant contribution to the scaling of $\Omega_n$ with respect to $\Phi$ in our system arises form a matrix element asymmetry $\beta_2$. This enables sub-harmonic transitions to appear at lower amplitudes compared to transmon qubits, which suggests the fluxonium to be a more suitable circuit for the application of sub-harmonic driving.

To confirm the predicted amplitude-dependent frequency shift as a function of drive amplitude, we measure the qubit frequency at a fixed drive amplitude and optimize the drive parameters using an iterative tune-up procedure as described in detail in Appendix~\ref{appendix:tune-up}. The measured amplitude-dependent frequency shift [Fig.~\ref{fig:fig2}\,(d), blue circles], increases to first order quadratically as predicted. 
Its curvature is, however, smaller, which we attribute to the frequency-dependent transfer function of the flux line caused by the skin effect in the stainless-steel coaxial wiring. The signal amplitude $\Phi_\mathrm{qubit}$ arriving at the qubit can be approximately characterized by an attenuation factor $a_0$ quantifying the length, thickness, and conductivity of the inner conductor $\Phi_\mathrm{qubit}=\Phi/(1+a_0\sqrt{\omega_d/2\pi})$~\cite{wigingtonTransientAnalysisCoaxial1957,heinrichQuasiTEMDescriptionMMIC1993}.
We achieve good agreement between model and data with an attenuation factor of $a_0=9.49(8)\times10^{-5}\,\mathrm{Hz}^{-1/2}$, resulting from a least-squares fit with $a_0$ as the single free parameter. 
The model predicts a power loss of $6.0(1)$\,dB at 1\,GHz, consistent with the specified total power loss of approximately 5.7\,dB at 1\,GHz for the signal line used in the experiment, as described in Appendix~\ref{appendix:setup}.
It is important to note, however, that the actual transfer function may be more complex due to additional distortions caused by dielectric losses and impedance mismatches in the flux line~\cite{rolTimedomainCharacterizationCorrection2020}. 

\section{\label{coherent_control}N-Photon Rabi Frequencies}  

\begin{figure}
    \centering
    \includegraphics{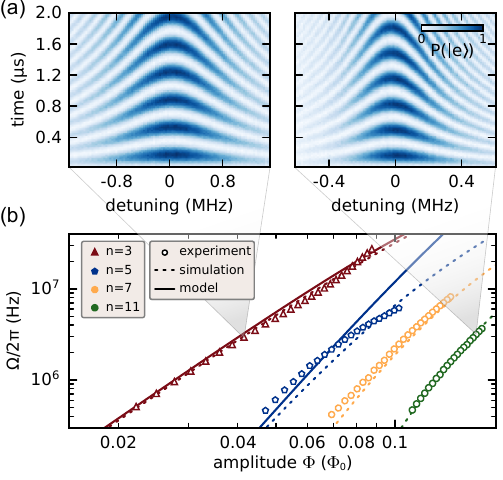}
    \caption{\textbf{Rabi frequencies for n-photon sub-harmonic drives.} (a) Excited-state population as a function of time and detuning for the 3$^{\rm rd}$ sub-harmonic at a drive amplitude of 41.5\,m$\Phi_0$ (left panel) and for the 11$^{\rm th}$ sub-harmonic at a drive amplitude of 160.5\,m$\Phi_0$ (right panel). The detuning denotes the offset of the drive from the calibrated center-frequency. The drive amplitudes are chosen to obtain similar Rabi frequencies of $\Omega/2\pi\sim 2.5$\,MHz. (b) Rabi frequencies $\Omega_n$ as a function of pulse amplitude $\Phi$ for different sub-harmonic drives, compared to numerical simulation and the analytic model. Extracted error bars are smaller than the marker size.}
    \label{fig:fig3}
\end{figure}

Following the characterization of the frequency shift, we quantify the Rabi frequency $\Omega_n$ of the sub-harmonics $n=3,5,7$ and 11 as a function of drive amplitude $\Phi$.
For a fixed drive amplitude, varying the drive frequency and pulse length reveals characteristic chevron patterns for each sub-harmonic drive, as shown exemplary for the 3-photon and 11-photon transitions in Fig.~\ref{fig:fig3}\,(a). 
We extract $\Omega_n$ for each amplitude using an iterative gate tune-up routine as described in Appendix \ref{appendix:tune-up}. 
The resulting amplitude dependence of the Rabi frequency [Fig.~\ref{fig:fig3}\,(b)] agrees well with the analytical no-free-parameter-model and numerical simulations for the 3$^{\rm rd}$ sub-harmonic when including the attenuation due to the skin-effect discussed above. 
We use the same value of $a_0$ for all the sub-harmonic modes, which we extracted from the frequency shift of the 3$^{\rm rd}$ sub-harmonic. 
The strong agreement between experiment, simulation and theory highlights the validity of our model including the skin-effect. 
Also for the 5$^{\rm th}$ sub-harmonic, a comparison of the analytic solution and experimental data shows good agreement at low drive powers, which demonstrates that the perturbative approach can be extended beyond the 3-photon processes. 
The 3-level approximation of the fluxonium remains valid for describing higher photon processes. 
However, a clear discrepancy emerges between the experimental data and the model at higher amplitudes for the 5$^{\rm th}$ sub-harmonic due to the limited order of the Magnus expansion.
In contrast, the numerical simulation matches the experimental data qualitatively up to the 11$^{\rm th}$ photon process.
We attribute residual deviations between experiments and simulations to a non-trivial transfer function for the flux line caused by standing waves due to impedance mismatches (see Appendix~\ref{appendix:power_scaling}).
After accounting for the influence of the skin effect, we fit the power scaling of each sub-harmonic with a leading-order exponent and find 2.81(1), 3.87(1), 4.73(1), and 5.78(1) for $n = 3, 5, 7$, and $11$, respectively. 
This differs from the intuitive expectation that an n-photon process scale with $\Phi^n$. 
In fact, the derivation in Appendix~\ref{appendix:theory} demonstrates that a $\Phi^n$ scaling is only valid for small driving amplitudes. 
At higher drive amplitudes, correction terms such as $\Phi^{n+2}$ become significant, indicating the presence of additional coupling paths. The number of these paths can grow exponentially with increasing $n$~\cite{huangTheoryMultiphotonProcesses2025}, thereby modifying the simple power-law dependence of Rabi rates and frequency shifts.

\section{Benchmarking Single-Qubit Gates}
\label{single_qubit_gates}
To benchmark the performance of sub-harmonic driving for single-qubit gates, we compare three different configurations: A 3$^{\rm rd}$ sub-harmonic drive through the LP-filtered flux line (LP), a conventional resonant drive through the unfiltered flux line (UF) and a conventional resonant drive via the MW line. 
We use the 3$^{\rm rd}$ sub-harmonic as it shows a higher Rabi frequency for similar drive power compared to sub-harmonics with higher photon numbers, resulting in faster gate speeds at equal power levels (Fig.~\ref{fig:fig3}).

For gate calibration, we set a constant pulse amplitude of a pulse with a flat-top Gaussian envelope and optimize frequency and time of the pulse. 
Additionally, we calibrate virtual Z rotations to correct for phase changes acquired due to the power-dependent frequency shift during the pulse (Fig.~\ref{fig:fig2}). 
A detailed description of the tune-up procedure is provided in Appendix~\ref{appendix:tune-up}.
Similarly, we calibrate the on-resonance gate implemented using a a DRAG pulse~\cite{gambettaAnalyticControlMethods2011}. 
All gates are set to an equal length of 64\,ns to ensure that qubit coherence affects all gates similarly.
\begin{figure}
    \centering
    \includegraphics[width=\linewidth]{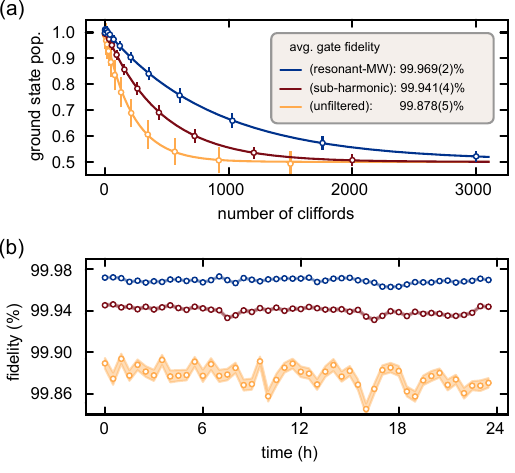}
    \caption{\textbf{Randomized Benchmarking of resonant and sub-harmonic single qubit gates.} (a) Ground state population as a function of number of random Clifford gates averaged over 24 hours. Each point is averaged over 1440 random sequences. (b) Time resolved average gate fidelity measured over 24 hours. Each data point shows the average gate fidelity obtained from 30 random sequences. The shaded region indicates one standard deviation.}
    \label{fig:fig4}
\end{figure}
We determine the gate fidelities of all three implementations through interleaved randomized benchmarking [Fig.~\ref{fig:fig4}\,(a)], where we use a gate set $\mathcal{G}=\{I, X_{\pm\pi}, Y_{\pm\pi}, X_{\pm\pi/2}, Y_{\pm\pi/2}\}$, resulting in 1.875 average gates per Clifford~\cite{barendsSuperconductingQuantumCircuits2014}.
To test the stability of the sub-harmonic gate to setup-related drifts that might affect the Rabi frequency $\Omega_n$ and subsequently the frequency shift $\delta_n$ ~\cite{lazarCalibrationDriveNonlinearity2023, xiaFastSuperconductingQubit2023}, we conduct repeated randomized benchmarking experiments over a time period of 24 hours.
The averaged and time resolved randomized benchmarking experiments shown in Fig.~\ref{fig:fig4}\,(a) and (b) result in gate fidelities $\mathcal F_{\mathrm{UF}}=$ 99.878(5)\,\% for the unfiltered flux line configuration (UF) and in the LP-filtered configuration in $\mathcal F_{\mathrm{SH}}=$ 99.941(4)\,\% for the sub-harmonic implementation, and $\mathcal F_{\mathrm{MW}}=$ 99.969(2)\,\% for the MW-line implementation.

Using the measured $T_1$ and $T_2^*$ times shown in Table \ref{coherence_table}, we compare the obtained gate fidelities to the coherence limit \cite{dingHighFidelityFrequencyFlexibleTwoQubit2023, pedersenFidelityQuantumOperations2007}
\begin{equation}
    \mathcal F^{\mathrm{coh}}=1-\frac{t_\mathrm{g}}{3}\left(\frac{1}{T_\phi}+\frac{1}{T_1}\right),
\end{equation}
which results in coherence limits of $\mathcal F_{\mathrm{UF}}^{\mathrm{coh}}=$ 99.87(2)\,\% and $\mathcal F_{\mathrm{LP}}^{\mathrm{coh}}=$ 99.965(4)\,\% for the unfiltered and LP-filtered flux line configuration, respectively, showing that both on-resonance gates are close to the coherence limit of the device. 
While the sub-harmonic gate shows a significantly improved fidelity compared to the unfiltered configuration, it does not reach the coherence limit.
Numerical simulations of the sub-harmonic gate show an infidelity due to leakage into non-computational states smaller than $10^{-5}$ (see Appendix~\ref{appendix:simulation}), suggesting that the observed infidelity is currently not limited by this effect. Instead, we identify three potential sources of infidelity arising from strong flux modulation, signal line distortions of the pulse and the dynamic frequency shift during the gate.
First, strong low-frequency flux modulations have been demonstrated in theory~\cite{huangEngineeringDynamicalSweet2021} and experiment~\cite{mundadaFloquetEngineeredEnhancementCoherence2020} to affect qubit coherence by creating Floquet quasienergy eigenstates, where the contribution of noise is averaged out over one modulation period.
As strong modulations enhance coherence only within a narrow range of drive amplitudes and frequencies, we consider it likely that the strong pulsed drive applied in our experiment leads to a reduction in qubit coherence not benefiting from the noise cancellation effect, thus negatively impacting gate performance.
Second, signal line distortion are a common problem detected in experiments~\cite{rolTimedomainCharacterizationCorrection2020}. 
In our experimental setup, we observe a signal echo 10\,ns after the pulse reducing the fidelity of consecutive gates.
To reduce this effect, we include a 24\,ns and 32\,ns delay subsequent to the $\pi$- and $\pi/2$-pulses within the 64\,ns gate duration.
To address this source of infidelity more directly, one can pre-distort the input pulse shape to account for the transfer-function of the flux line after characterizing the step-response with a cryoscope experiment~\cite{rolTimedomainCharacterizationCorrection2020}. 
Third, the gate fidelity could potentially be improved by accounting for the dynamic frequency shift during the gate operation. 
During the rise time of the flat-top Gaussian pulse, the qubit frequency shifts gradually, causing the drive frequency $\omega_\mathrm d$ to become resonant with the sub-harmonic transition only after the rise is complete. By tracking this frequency shift, the drive could remain on resonance throughout the entire gate duration, thereby improving both gate speed and fidelity.

We assess the stability of the different gate implementations by the variance of the gate fidelity shown in Fig.~\ref{fig:fig4}\,(b), where the average gate fidelity is obtained from 30 random Clifford sequences for 24 hours with a cadence of 30 minutes.
Using the variance of the MW-gate as a reference, we calculate the relative variance of the other gates to be $\sigma^2_\mathrm{SH}/\sigma^2_\mathrm{MW}=2.3$ for the sub-harmonic gate and $\sigma^2_\mathrm{UF}/\sigma^2_\mathrm{MW}=19.9$, showing that the sub-harmonic gate exhibits almost equal stability compared to the on-resonance MW-gate and close to one order of magnitude improvement to the resonant flux line implementation without filter, which is more sensitive to temperature fluctuations due to the $T_1\propto1/(2n_\mathrm{th}+1)$ dependence~\cite{blaisCircuitQuantumElectrodynamics2021}.

\section{Conclusion and Outlook}

In conclusion, we have demonstrated that a single low-pass filtered flux-control line is sufficient to realize universal, high-fidelity control of a fluxonium qubit via sub-harmonically driven multi-photon transitions, while simultaneously suppressing relaxation and dephasing due to control-line noise.
We have developed a theoretical model for sub-harmonic driving in strongly anharmonic systems that accurately predicts drive-induced frequency shifts and Rabi frequencies. The model reveals enhancement of the sub-harmonic Rabi frequency arising from matrix element asymmetries, emphasizing that sub-harmonic control is intrinsically more effective in fluxonium qubits than in weakly anharmonic systems like transmons.
Experimentally, we have demonstrated qubit control ranging from 3- to 11-photon processes, confirming that sub-harmonic control is particularly effective in fluxonium qubits.
Using optimized flat-top Gaussian pulses leads to gate fidelities exceeding 99.94\,\% approaching the fidelities achieved via conventional on-resonant driving which are close to the coherence limit of the device.
By further optimization using closed-loop schemes~\cite{werninghausLeakageReductionFast2021, kochQuantumOptimalControl2022} and pre-distorted pulses to correct for signal distortions in the coaxial cable~\cite{rolTimedomainCharacterizationCorrection2020}, we expect that sub-harmonic gates will perform equally to on-resonance gates in the near future. 

The on-chip integration of reflective low-pass filters will mitigate the remaining decay attributed to losses in the flux line segment that connects the chip to the filter. 
Moreover, the strong amplitude dependence of multi-photon transitions offers increased flexibility in choosing the drive frequency, which can be used to mitigate crosstalk~\cite{brinkDeviceChallengesTerm2018}.
Off-resonant driving of neighboring qubits is further reduced due to the higher-order amplitude dependence of the Rabi frequency compared to on-resonance driving.

By combining DC flux biasing and high-fidelity transversal control into a single Purcell-protected channel, our approach establishes a scalable and hardware-efficient control architecture for fluxonium-based processors, matching fixed-frequency transmons in terms of wiring simplicity.
Beyond fluxonium qubits, sub-harmonic control exploits the presence of higher-lying energy levels to enable transitions that are otherwise forbidden by selection rules or suppressed by frequency filtering, provided appropriate matrix elements exist. This makes it a promising strategy for controlling other strongly anharmonic qubit circuits, such as qudits and protected qubits where transitions between computational states are exponentially suppressed but transitions to higher levels remain accessible.

\section*{Acknowledgements}
We thank Jacquelin Luneau and Peter Rabl for insightful discussions and helpful comments.  This work received financial support from  the German Federal Ministry of Education and Research via the funding program ’Quantum technologies - from basic research to the market’ under contract number 13N15680 (GeQCoS) and under contract number 13N16188 (MUNIQC-SC), by the Deutsche Forschungsgemeinschaft (DFG, German Research Foundation) via project number FI2549/1-1 and via the Germany’s Excellence Strategy EXC-2111-390814868 ‘MCQST’ as well as by the European Union by the EU Flagship on Quantum Technology HORIZON-CL4-2022-QUANTUM-01-SGA project 101113946 OpenSuperQPlus100. The research is part of the Munich Quantum Valley, which is supported by the Bavarian state government with funds from the Hightech Agenda Bayern Plus.

\section*{Data Availability}
The data that support the findings of this paper are openly available~\cite{schirk_2025_15841399}.

\appendix
\section{\label{appendix:qubit-params}Device Parameters}
\begin{figure}
    \centering
    \includegraphics{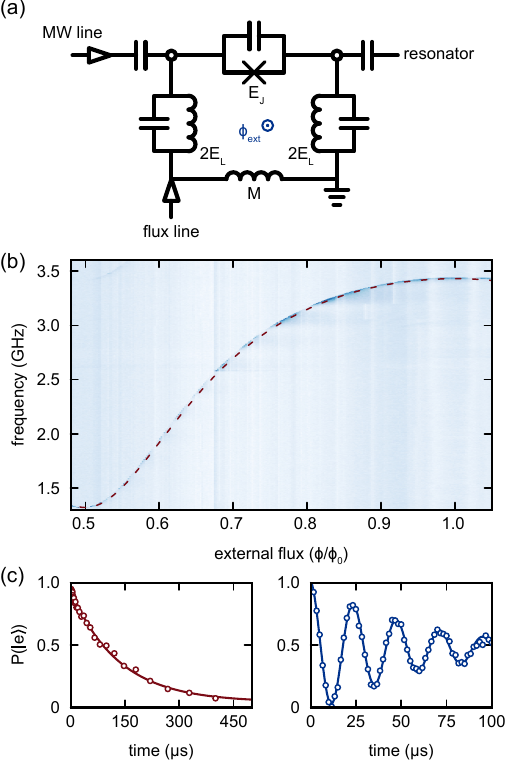}
    \caption{\textbf{Device Characterization}. (a) Equivalent circuit of the fluxonium qubit, realized with a Josephson junction with energy $E_\mathrm{J}$, a split inductance with inductive energy $E_\mathrm{L}$ and a total capacitive energy $E_\mathrm{C}$. (b) Pulsed spectroscopy of the fundamental fluxonium mode as a function of external flux. The red line represents the transition frequency obtained from circuit simulations. (c) Exemplary decay (left panel) and coherence (right panel) measurements of the device using sub-harmonic driving.}
    \label{fig:spectrum}
\end{figure}
Fig.~\ref{fig:spectrum}\,(a) shows a lumped element circuit of the qubit used in this work including characterization measurements for the device in (b) and (c). We realize the qubit with a niobium ground plane on a high resistivity silicon substrate, and with Al/AlO$_x$/Al Josephson junctions both for phase-slip junction and the inductance. The inductance is split into two arrays, allowing for galvanic connection of the flux line~\cite{moskalenkoTunableCouplingScheme2021, moskalenkoHighFidelityTwoqubit2022}. The coupling of the flux line to the qubit is given by their shared inductance $M$. Both the readout resonator as well as the MW drive line are coupled capacitively, predominantly to one of the two qubit islands.
From qubit spectroscopy of the fundamental mode, depicted in Fig.~\ref{fig:spectrum}\,(b), and comparing to eigenenergy simulations of the equivalent circuit obtained using the ScQubits package~\cite{groszkowskiScqubitsPythonPackage2021,chittaComputeraidedQuantizationNumerical2022}, we determine the circuit parameters $E_\mathrm{J}/h$=1.69\,GHz, $E_\mathrm{L}/h=1.07$\,GHz and $E_\mathrm{C}/h=0.68$\,GHz, resulting in a qubit frequency $\omega_\mathrm{eg}/2\pi$=1.32\,GHz and an anharmonicity $\alpha/2\pi=0.81$\,GHz. The mutual inductance value $M=3.4$\,pH is determined following a calculation given in Appendix~\ref{appendix:temperatures}. Fig.~\ref{fig:spectrum}\,(c) shows exemplary single trace measurements for the energy relaxation time $T_1$ (left) and Ramsey-coherence time $T_2^*$ of the device. For a more detailed analysis we refer the reader to Section~\ref{protection} in the main text.

\section{\label{appendix:setup}Experimental Setup}
\begin{figure}
    \centering
    \includegraphics[width=\linewidth]{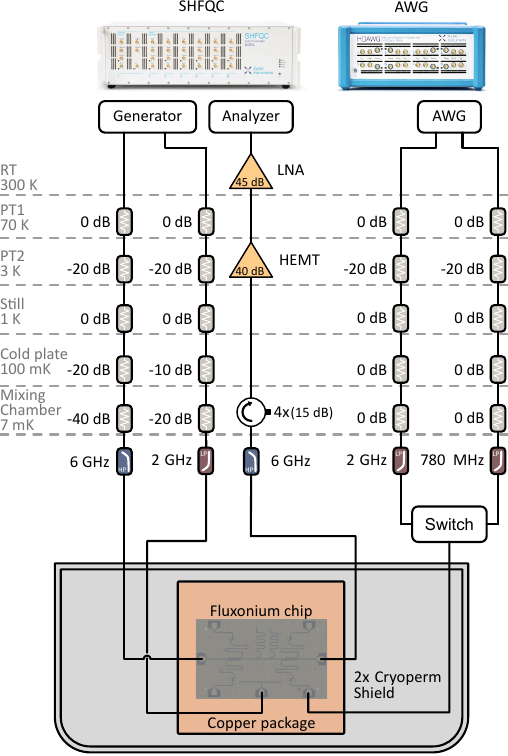}
    \caption{\textbf{Experimental Setup.} The qubit chip is mounted inside two cryoperm shields to protect it from stray magnetic fields (For details see main text).}
    \label{fig:setup_full}
\end{figure}

The sample is mounted at the mixing chamber stage of a dilution refrigerator (\textit{Bluefors} XLD1000sl). 
Fig.~\ref{fig:setup_full} displays the full electronic setup up to the room temperature control.
We employ two devices for qubit control.
A \textit{Zurich Instruments} quantum controller (SHFQC) is used for driving the qubit at microwave frequencies and readout of the qubit state, while a single channel of a \textit{Zurich Instruments} arbitrary waveform generator (HDAWG) is utilized for DC-flux biasing and AC-flux control.
50~dB of attenuation combined with a 2~GHz low pass filter (\textit{KL} 6L250-2000) ensure a low noise MW-control control line.
The SHFQC generates and digitizes the readout signal.
With a HEMT (\textit{LNF}-LNC4\_8C) and a room temperature amplifier (\textit{Qotana} DBLNA104000800F) we achieve a state assignment fidelity of $F_{\mathrm{readout}}\approx94\,\%$.
We use a high-pass filter (\textit{Minicircuits} VHF-5050+) at the input and output ports of the readout line in combination with two dual-junction isolators (\textit{LNF}-ISISC4\_12A) at the output to protect the qubits from unwanted noise photons in resonators.
Flux control is achieved with a single line that is  attenuted by 20~dB at the 3~K stage. Through a switch (\textit{Radiall} R573423600) we test different filter configurations at the 7~mK plate.
The chip is packaged and mounted inside two cryoperm shields. 
To estimate the losses of the flux line we characterize the 3\,m room temperature coaxial cable using a vector network analyzer (VNA), resulting in 2.5\,dB of attenuation at 1\,GHz. Inside the cryostat we assume a total length of 1.05\,m with attenuation of around 3\,dB at 1\,GHz extracted from the specification sheet of Bluefors~\cite{krinnerEngineeringCryogenicSetups2019}. This results in a total power loss of around $-5.7$\,dB at $1$\,GHz, which agrees well with the value of -6.0(1)\,dB predicted from the skin-effect described in Section~\ref{parametic_chapter}. 

\section{$\mathrm{T_2}$-limit}\label{appendix:T2Limit}
There are two main sources of noise that lead to dephasing and subsequently to a reduction of the coherence time~\cite{krantzQuantumEngineerGuide2019}. On the one hand, low frequency noise that exhibits the characteristic 1/f behavior in the power spectral density over the frequency is present in the system~\cite{yanRotatingframeRelaxationNoise2013}. On the other hand, fluctuations of the photon number in the readout resonator dispersively shift the qubit's frequency and lead to dephasing of the qubit. We can express the effect of thermal resonator photons $n_\mathrm{th} < 1$ on the qubit dephasing time $ T_{\varphi}$ as~\cite{clerkUsingQubitMeasure2007}
\begin{equation}
    \frac{1}{T_{\varphi}}=\frac{\kappa\left(2\chi\right)^2}{\kappa^2 + \left(2\chi\right)^2}n_\mathrm{th}
    \label{dephasing_rate}
\end{equation}

Here $\kappa$ is the resonator linewidth, $\chi$ the dispersive shift and $n_\mathrm{th}$ the average thermal resonator population. 
The dephasing rate decreasing the maximum coherence time as
\begin{equation}
    \frac{1}{T_2}=\frac{1}{2T_1}+\frac{1}{T_{\varphi}}.
    \label{rates}
\end{equation}
By measuring $T_1$ and $T_2$,and assuming that the fluxonium is first-order protected from flux noise at its sweet spot, we extract the resonator population and calculate an effective resonator temperature using Eq.~\ref{dephasing_rate} and Eq.~\ref{rates}. For our experimental parameters $\kappa/2\pi=1.2$~MHz, $2\chi/2\pi = 5.3$\,MHz, $\omega_R/2\pi=6.9$\,GHz, $\mathrm{T_2}=75(9)$\,\textmu s and $T_1=168(20)$\,\textmu s (see Table~\ref{coherence_table}), resulting in an effective resonator temperature of 51(1)\,mK.

\section{\label{appendix:temperatures}Qubit temperature and bath coupling}
\begin{figure}
    \centering
    \includegraphics{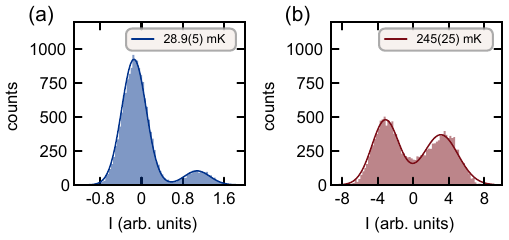}
    \caption{\textbf{Thermal state measurements of different filter configurations}. Thermal state projected onto the state discrimination axis. (a) filtered configuration. (b) un-filtered configuration.}
    \label{fig:temperature}
\end{figure}

We extract the effective qubit temperature from single shot measurements of the qubit's thermal population, shown in Fig.~\ref{fig:temperature}. The qubit temperature is approximated by~\cite{pathriaStatisticalMechanics2011}
\begin{equation}
    T_\mathrm{eff}=-\frac{\hbar\omega}{k_\mathrm{B}} \ln\left(\frac{P(|e\rangle)}{P(|g\rangle)}\right)^{-1},
\end{equation}
where $\hbar$ is the reduced Planck's constant, $k_\mathrm{B}$ is the Boltzmann's constant and $P(|e\rangle)$ ($P(|g\rangle)$) the probability for finding the qubit in the ground state $|g\rangle$ (excited state $|e\rangle$), respectively. 
Fig. \ref{fig:temperature} shows the measured ground state population $P(|g\rangle)$ for the filtered (a) and unfiltered (b) configuration. By fitting the data with a double Gaussian and extracting the relative magnitude of the peak heights, we extract effective qubit temperatures of $T=28.9(5)$\,mK and $T=245(25)$\,mK for the filtered and unfiltered case, respectively.
In the following, we provide an estimate for the qubit-bath coupling via the flux line. The average thermal bath photon number at a frequency $\nu$ and temperature $T$ is given by the Bose-Einstein statistic:
\begin{equation}
    n_\mathrm{th}=\left[\exp\left({\frac{\hbar\omega}{k_\mathrm{B}T}}\right)-1\right]^{-1}.
    \label{eq:bose-einstein}
\end{equation}
We calculate the decay rate caused by baths $i$ coupled to the qubit with a coupling rate $\gamma_i$ by~\cite{blaisCircuitQuantumElectrodynamics2021}
\begin{equation}
    \gamma(T) = \sum_i\gamma_{i}[2n_{\mathrm{th},i}(T)+1].
    \label{eq:thermal_decay}
\end{equation}
By adding a filter on the line, we reduce the coupling rate to the bath by the attenuation of that filter. Thus we can extract the initial bath coupling rate $\gamma_0$ by taking the difference of decay rates between the filtered (LP) and unfiltered case (UF):
\begin{equation}
    (1/T_{1\mathrm{,LP}}-1/T_{1\mathrm{,UF}})=\gamma_0(A-1)(2n_\mathrm{th}+1),
\end{equation}
where $A$ is the power attenuation factor of the filter at the qubit frequency and $\gamma=1/T_1$. Using experimental values stated in Section~\ref{protection}, an attenuation factor $A=10^{-3.55}$ of the filter at the qubit frequency and a bath temperature of 3\,K, we estimate a decay time into this channel of $1/\gamma_0$=3.6(5)\,ms. 
From this result, we can extract the mutual inductance $M$ between the qubit and the flux line. 
The equation relating the $\gamma_\mathrm{eg}$ and $M$ can be derived from Fermi's golden rule, following~\cite{kochChargeinsensitiveQubitDesign2007,popCoherentSuppressionElectromagnetic2014a}:
\begin{equation}
    \gamma_\mathrm{eg}=\frac{\omega_\mathrm{eg}}{2\pi}\frac{R_\mathrm{Q}R}{|Z|^2}\frac{M^2}{L^2}|\langle \mathrm g|\hat{\varphi}| \mathrm e\rangle|^2\left[1+\coth\left(\frac{\hbar\omega_\mathrm{eg}}{2k_\mathrm{B}T}\right)\right],
    \label{eq:fl_decay}
\end{equation}
where $Z$ is the impedance of the flux line, $L$ the inductance of the qubit and $R_\mathrm{Q}=h/(2e)^2$ the superconducting resistance quantum. Using the qubit parameters from Appendix \ref{appendix:qubit-params} and solving Eq.~(\ref{eq:fl_decay}) for $M$, we obtain $M_\mathrm{Exp}=3.1(2)$\,pH in the limit of $T=0$, which is close to the design value of 3.2\,pH simulated using 3D-MLSI~\cite{khapaev3DMLSISoftwarePackage2001}. 
Equivalently, using $M_\mathrm{Design}$ in Eq.~(\ref{eq:fl_decay}) gives $\gamma_0=1/3.4$\,ms.
We simulate the relevant matrix elements for our circuit using the ScQubits package~\cite{groszkowskiScqubitsPythonPackage2021,chittaComputeraidedQuantizationNumerical2022}.

\section{Numerical Simulations\label{appendix:simulation}}
\begin{figure}
        \centering
        \includegraphics{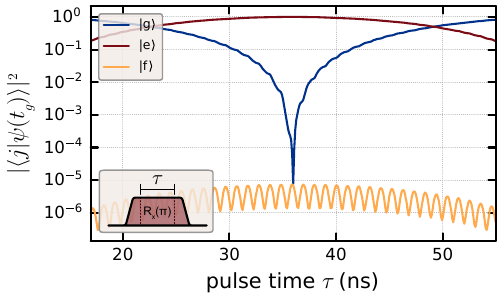}
        \caption{\textbf{Simulated state overlap vs pulse duration}. We simulate the time evolution of the fluxonium qubit driven at $\omega_\mathrm{d} = 480$\,MHz with a drive amplitude of $\Phi = 5.25$\,m$\Phi_0$, while varying the duration $\tau$ of the flat section of a flat-top Gaussian pulse. The state populations of the computational states $|g\rangle$, $|e\rangle$, and $|f\rangle$ are shown as a function of $\tau$. The inset illustrates the relation between $\tau$ and the overall pulse shape. The total rise and fall times of the pulse sum to 16\,ns.
}
        \label{fig:sim_vs_time}
    \end{figure}

    To extract the position of sub-harmonic drive frequencies $\omega_\mathrm d$ and their respective Rabi frequencies $\Omega$ at different drive amplitudes $\Phi$, we use a discretized approach for the simulation of the time dynamics of our system. We start with the Hamiltonian with an applied static flux bias of $0.5\,\Phi_0$
    \begin{equation}
        \hat{H}(t) = 4E_\mathrm{C}\hat{n}^2 + E_\mathrm{J}\cos{\hat{\varphi}} + \frac{E_\mathrm{L}}{2}[\hat{\varphi} - \phi(t)]^2.
    \end{equation}
    We then diagonalize numerically for zero applied flux, yielding the eigenstates $|\psi_i\rangle$, which we truncate to the 20 lowest energy eigenstates to reduce computation time, while maintaining very good numerical precision. We then discretize the pulse shape
    \begin{equation}
        \phi(t_j)=2\pi\frac{\Phi}{\Phi_0}E(t_j)\cos(\omega_\mathrm{d}t_j),
    \end{equation}
    where $E(t)$ is the pulse envelope and $\omega_\mathrm{d}$ the drive frequency. We use a time resolution of $t_{j+1}-t_j=0.1$\,ns such that the oscillation frequency of the drive is slow compared to the time interval. The time evolution of the system is simulated by iteratively applying the time evolution operator. For each time slice $t_j$, the state after discretized time evolution is given by
    \begin{equation}
        |\psi(t_{j+1})\rangle = \exp\left(-i\frac{\hat{H}(t_{j+1})t_j}{\hbar}\right)|\psi(t_j)\rangle,
    \end{equation}
    with the initial condition $|\psi(0)\rangle=|g\rangle$. After completion of the time evolution at the pulse length $t_\mathrm{pulse}$, we extract the excited state population from the overlap $\langle e|\psi(t_\mathrm{pulse})\rangle$.
    We find the transitions corresponding to sub-harmonic processes with an iterative optimization. 
    For a fixed drive amplitude, we perform an interleaved optimization of the drive frequency and pulse time for applying an on-resonance $\pi$-rotation, optimizing one parameter while keeping the other fixed. \\
    To assess the coherent error of the sub-harmonic single-qubit gate shown in Section~\ref{single_qubit_gates}, we simulate the time-evolution at $\omega_\mathrm d=480$\,MHz, corresponding to a drive amplitude $\Phi=5.25$\,m$\Phi_0$ for different pulse lengths, as shown in Fig.~\ref{fig:sim_vs_time}.
    At the optimal pulse time for a $\pi$-pulse, the fidelity is limited by leakage into the $|f\rangle$-state to 7.4$\times10^{-6}$, which is well below the experimentally observed gate fidelity.
\section{Model}
\label{appendix:theory}
In this section, we derive an analytical model describing the sub-harmonic driving of a fluxonium qubit at its 3$^{\rm{rd}}$ sub-harmonic. Higher-order harmonics can be performed analogously. The analytic expressions for $5^{\rm th}$ sub-harmonics are used in Fig.~\ref{fig:fig3}, even though not explicitly stated here.\\
A flux driven fluxonium, biased at flux sweet-spot $0.5\Phi_0$, is described by the following Hamiltonian ($\hbar=1$)~\cite{mundadaFloquetEngineeredEnhancementCoherence2020}
\begin{equation}
    \hat{H}(t) = 4E_\mathrm{C}\hat{n}^2 + E_\mathrm{J}\cos{\hat{\varphi}} + \frac{E_\mathrm{L}}{2}\left[\hat{\varphi} - \phi(t)\right]^2,
    \label{app:hamiltonian}
\end{equation}
with parameter definitions equal to Eq. (\ref{eq:hamiltonian}). 
By expanding the square, we can remove a time-dependent, global energy-shift from the system and find that the resulting drive-part of the Hamiltonian actually is linear in $\phi(t)$.\\
By diagonalizing the rest-frame Hamiltonian $\hat H_0=\hat H|_{\phi=0}$ and utilizing the ladder operators, $\hat b$ and $\hat{b}^\dagger$ in the energy basis of the fluxonium, i.e. $\hat b|n\rangle=\sqrt{n}|n-1\rangle$, the system takes the form:
\begin{align}
    \hat H(t) &= \hat H_0 - \phi(t)\; E_\mathrm{L} \hat \varphi\\
    \hat H_0 &= \omega_\mathrm{eg} \hat b^\dagger \hat b + \frac{\alpha}{2} \hat b^\dagger \hat b^\dagger \hat b\hat b + \cdots \; ,
\end{align}
with the qubit frequency $\omega_\mathrm{eg}$, its anharmonicity $\alpha$. The terms proportional to higher powers of $\hat b$ are suppressed for the moment and are sub-relevant for the Rabi oscillation.\\
The next step is to express the drive-term linear in $\hat \varphi$ in the same basis. For that, we realize that $\hat H_0$ in its formulation (\ref{app:hamiltonian}), is symmetric in $\hat \varphi$, hence its eigenfunctions $|m\rangle$ need to be either symmetric or asymmetric in $\hat \varphi$. Therefore, the matrix elements $\langle m| \hat \varphi|m'\rangle$ can only be non-vanishing if $m-m'$ is odd. Given that in our application the matrix elements are sufficiently small for $m-m'\leq 3$, the driven part of the Hamiltonian can be fully characterized by $\sum_n^m \beta_n= E_\mathrm{L} \langle m-1|\hat\varphi|m\rangle / \sqrt{m}$. Without loss of generality, we can therefore express
\begin{align}
    E_\mathrm{L}\hat\varphi = \beta_1 (\hat b+\hat b^\dagger) + \beta_2 (\hat b^\dagger\hat b\hat b +\hat b^\dagger\hat b^\dagger\hat b) +\cdots\; ,
\end{align}
We note again that terms with powers of $\hat b$ higher than three exist but are not relevant to accurately describe Rabi oscillations.\\
We are interested in the effective dynamics of the model, given by
\begin{align}
    \hat U(0,t_\mathrm{pulse}) = \mathcal{T}\exp\left(-i \int_0^{t_\mathrm{pulse}} dt \hat H(t) \right),
\end{align}
denoting the time-ordered exponential.
To study the $n^{\rm{th}}$ sub-harmonic drive, it is useful to go into a frame which rotates at $n$-times the driving frequency, $\omega_\mathrm{d}$. 
The frame change is achieved by the unitary transformation $\hat R(t)=\exp(it n\omega_\mathrm{d}\hat b^\dagger \hat b)$.
Further, to work with unit-less quantities, we rescale the rotating system by $t\mapsto \tau= 2t\omega_\mathrm{d}$ and $\hat H(t)\mapsto \hat{\tilde {H}}(\tau):=(\hat R\hat H(t)\hat R^\dagger-\hat R\dot{\hat R}^\dagger)/(2\omega_\mathrm{d})$. 
Here, we are focusing on the 3$^{\rm rd}$ sub-harmonic and  thus express the Hamiltonian in a rescaled three-photon frame as:
\begin{widetext}
\begin{align}
    \hat{\tilde {H}}(\tau) = \frac{\Delta}{2\omega_\mathrm{d}}\hat b^\dagger \hat b +\frac{\alpha}{4\omega_\mathrm{d}}\hat b^\dagger \hat b^\dagger \hat b \hat b - \frac{\bar \phi}{2\omega_\mathrm{d}}E(t)\left((e^{i\tau}+e^{i2\tau})(\beta_1 \hat b + \beta_2 \hat b^\dagger\hat b\hat b)+h.c.\right)+\cdots\;,
    \label{app:r-3p-Ham}
\end{align}
\end{widetext}
where we denote the detuning between qubit frequency and triple drive frequency by $\Delta= \omega_\mathrm{eg}-3\omega_\mathrm{d}$ and use $\phi(t)=\bar{\phi}E(t)\cos(\omega_\mathrm{d}t)$, with $E(t)$ being an envelope function with unit amplitude.\\
Working in the dispersive regime, we assume the following quantities to be small
\begin{align}
    \Delta/(2\omega_\mathrm{d}) \ll 1,\quad \alpha/(4\omega_\mathrm{d}) \ll 1,\quad \beta_m \bar \phi /(2\omega_\mathrm{d}) \ll 1
    \label{app:perturb}
\end{align}
In the regime where (\ref{app:perturb}) holds, one expects that the time evolution can be well approximated by a power series with respect to those small quantities with convergence rate given by (\ref{app:perturb}). Hence, we denote $\hat{\tilde {H}} \equiv \mathcal{O}(\omega_\mathrm{d}^{-1})$. We want to stress at this point that the device is not in the low-frequency fluxonium regime and $\alpha/(4\omega_\mathrm{d})\sim 0.5$. For heavy fluxonia $\alpha\gg \omega_\mathrm{d}$, the approximation by a power series in (\ref{app:perturb}) does not hold anymore. It is possible to reformulate the equations to account for general $\alpha/\omega_\mathrm{d}$, however this requires a more careful investigation beyond the scope of this paper~\cite{huangTheoryMultiphotonProcesses2025}.

A well-known procedure to perturbatively obtain the finite-time evolution operator from a time-dependent Hamiltonian is the Magnus expansion \cite{magnusExponentialSolutionDifferential1954}. Further, for periodically driven system, the extended Floquet-Magnus expansions can be used to obtain an effective Hamiltonian, which is already implemented in time-modulated cold-atom systems~\cite{goldmanPeriodicallyDrivenQuantum2015}. 

The aim of Floquet-Magnus expansions is to move into a frame in which the Hamiltonian becomes independent of the fast-rotating contributions, i.e. we aim for a unitary transformation $\hat U(\tau) = \exp(i \hat K(\tau) )$ such that
\begin{align}
    \hat H_{\rm eff} = \hat U \hat{\tilde {H}} \hat U^\dagger - i \hat U \partial_{\tau} \hat U^\dagger
    \label{app:time_ind_frame}
\end{align}
becomes independent of terms of the form $\exp(ik\tau)$. Note that the time-dependence of the envelope function $E(t)$ remains, which can be assumed to be slow since the time length of $\dot{E}(t)$ is negligible compared to pulse time. To eliminate rapidly oscillating terms, we generate $\hat K(\tau)$ iteratively by powers of (\ref{app:perturb}), i.e. we define:
\begin{align}
    \begin{aligned}
        \hat{\tilde {H}} = \sum_k e^{ik\tau} \hat H_k,\quad \hat H_k&\equiv \mathcal{O}(\omega_\mathrm{d}^{-1}),\\
        \hat H_{\rm eff} =\sum_n \hat H_{\rm eff}^{(n)},\quad \hat H_{\rm eff}^{(n)}&\equiv \mathcal{O}(\omega_\mathrm{d}^{-n}),\\
        \hat K(\tau) =\sum_{n}K^{(n)},\quad \hat K^{(n)}&=\sum_{k\neq 0}e^{ik\tau} \hat K_{k}^{(n)},\\
        &\hat K_{k}^{(n)}\equiv \mathcal{O}(\omega_\mathrm{d}^{-n})\;.
    \end{aligned}
\end{align}
Expanding (\ref{app:time_ind_frame}) in powers of $1/\omega_\mathrm{d}$, we obtain the defining equations for $\hat H_{\rm eff}^{(n)}$ and $\hat K_{k}^{(n)}$. At first order, it reads
\begin{align}
    \hat H^{(1)}_{\rm eff}&= \hat{\tilde {H}}(\tau) -\partial_\tau \hat K^{(1)}
\end{align}
and by defining $\hat K^{(n)}$ such that it collects all $\tau$-dependency inside it, the Hamiltonian becomes effectively time-independent,
\begin{align}
     \hat H^{(1)}_{\rm eff}=\hat H_0,\quad \hat K_k^{(1)} = \frac{\hat H_k}{ik}
\end{align}
One proceeds similarly for higher orders:
\begin{align}
    \hat H^{(2)}_{\rm eff} &=[i \hat K^{(1)},\hat{\tilde {H}}]-\frac{1}{2}[i\hat K^{(1)},\partial_\tau \hat K^{(1)}]-\partial_\tau \hat K^{(2)}\nonumber\\
    &\begin{aligned}
        \Leftrightarrow \hat H^{(2)}_{\rm eff}=&\sum_{k\neq0}\frac{[\hat H_k,\hat H_{-k}]}{2k},\\
        K_k^{(2)} = &\frac{[\hat H_k,\hat H_0]}{ik^2}+\sum_{k'\neq0}\frac{[\hat H_{k'},\hat H_{k-k'}]}{2ik'k}
    \end{aligned} \\ \nonumber
\end{align}
\begin{align}
    \hat H^{(3)}_{\rm eff} &=[i \hat K^{(2)}(\tau), \hat{\tilde {H}}(\tau)]+ \frac{1}{2}[i\hat K^{(1)}(\tau), [i\hat K^{(1)}(\tau), \hat{\tilde {H}}(\tau)]] \nonumber\\&- \frac{1}{2}[i \hat K^{(2)}(\tau), \partial_\tau \hat K^{(1)}(\tau)]- \frac{1}{2}[i \hat K^{(1)}(\tau), \partial_\tau \hat K^{(2)}(\tau)] \nonumber\\
    &- \frac{1}{6}[i \hat K^{(1)}(\tau), [i\hat K^{(1)}(\tau),\partial_\tau \hat K^{(1)}(\tau)]] -\partial_\tau \hat K^{(3)}(\tau)\nonumber\\
    &\begin{aligned}
        \Leftrightarrow & \hat H^{(3)}_{\rm eff}= \sum_{k\neq0}\frac{[[\hat H_k,\hat H_0],\hat H_{-k}]}{2k^2} \\
        &+\sum_{\substack{k,k'\neq0 \\ k-k'\neq0}}\frac{[[\hat H_{k'},\hat H_{k-k'}],\hat H_{-k}]}{4k}\left(\frac{1}{k'}+\frac{1}{3(k'-k)}\right)
    \end{aligned}\nonumber\\
    &\cdots\label{eq:order_expansion}
\end{align}
Note, that we omit the definition of $\hat K^{(3)}$ as it is not relevant in order to obtain the effective Hamiltonian at third order.

It can be shown that $\hat K^{(n)}$ always depends on the envelope function, such that $\exp(i\hat K(\tau=0))=\exp(i\hat K(\tau=2\omega_\mathrm{d}t_\mathrm{pulse}))=\mathds{1}$ for an envelope function which vanishes at the beginning and the end of the pulse, e.g. a flat-top Gaussian. Therefore, the effective Hamiltonian captures the dynamics of the full gate.\\
Explicitly for the three-photon process described by $\hat{\tilde {H}}$ in (\ref{app:r-3p-Ham}), we obtain the effective Hamiltonian: 
\begin{align}\label{eq:effectiveH_1}
    &\hat H^{(1)}_{\rm eff,3} = \Delta \hat b^\dagger \hat b + \frac{\alpha}{2} \hat b^\dagger \hat b^\dagger \hat b\hat b\\ 
    &\hat H^{(2)}_{\rm eff,3} = -\frac{3\beta_2(2\beta_1+\beta_2)}{8\omega_\mathrm{d}}E^2\bar{\phi}^2 \hat b^\dagger \hat b
    \label{eq:effectiveH_2}
\end{align}
\begin{widetext}
\begin{align}
    &\begin{aligned}
        \hat H^{(3)}_{\rm eff,3} = &\frac{5(\alpha (\beta_1+\beta_2)^2+\beta_2(2\beta_1+\beta_2)\Delta)}{32\omega_\mathrm{d}^2} E^2\bar{\phi}^2 \hat b^\dagger \hat b   +\frac{\beta_1\beta_2(2\beta_1+\beta_2)}{32\omega_\mathrm{d}^2}E^3\bar{\phi}^3(\hat b^\dagger + \hat b) 
        \end{aligned} \\
&\begin{aligned}
        \hat H^{(4)}_{\rm eff,3} = &-\frac{21\beta_2^2(7\beta_1^2+10\beta_1\beta_2+4\beta_2^2)}{512\omega_\mathrm{d}^3} E^4\bar{\phi}^4 \hat b^\dagger \hat b 
        -\frac{9(\alpha (\beta_1+\beta_2)^2(\alpha+2\Delta)+\beta_2(2\beta_1+\beta_2)\Delta^2)}{128\omega_\mathrm{d}^3} E^2\bar{\phi}^2 \hat b^\dagger \hat b\\
        &+\frac{\beta_1(6\alpha(\beta_1+\beta_2)^2+13\beta_2(2\beta_1+\beta_2)\Delta)}{768\omega_\mathrm{d}^3}E^3\bar{\phi}^3(\hat b^\dagger + \hat b)
    \end{aligned}\\
    &\begin{aligned}
        \hat H^{(5)}_{\rm eff,3} =&\frac{\alpha\beta_2(313\beta_1+527\beta_2)(\beta_1+\beta_2)^2}{2048\omega_\mathrm{d}^4} E^4\bar{\phi}^4 \hat b^\dagger \hat b +\frac{107\beta_2^2(7\beta_1^2+10\beta_1 \beta_2+4\beta_2^2)\Delta}{2048\omega_\mathrm{d}^4} E^4\bar{\phi}^4 \hat b^\dagger \hat b \\
        &+\frac{17(\alpha(\beta_1+\beta_2)^2(\alpha^2+3\alpha\Delta+3\Delta^2)+\beta_2(2\beta_1+\beta_2)\Delta^3)}{512\omega_\mathrm{d}^4} E^2\bar{\phi}^2 \hat b^\dagger \hat b \\
        &-\left(\frac{\beta_1(29\beta_2(2\beta_1+\beta_2)\Delta^2+6\alpha(\beta_1+\beta_2)^2(\alpha+3\Delta))}{3072\omega_\mathrm{d}^4}E^3\bar{\phi}^3 + \frac{\beta_1\beta_2^2(29\beta_1^2+38\beta_1\beta_2+14\beta_2^2)}{1024\omega_\mathrm{d}^4}E^5\bar{\phi}^5\right)(\hat b^\dagger + \hat b)
        \label{eq:effectiveH_5}
    \end{aligned}
\end{align}
\end{widetext}
Computing the effective Hamiltonian up to the $5^{\rm th}$ order provides a good match with the experimental data of the device. If one would delve deeper into the fluxonium regime, i.e. increasing the ratios in (\ref{app:perturb}) of the relevant quantities versus $\omega_\mathrm{eg}$, higher orders of the power series will be necessary. On the other hand, high-frequency devices (such as transmon qubits) are found to have very low ratios and will in general require much lower orders to be captured accurately (e.g. enabling tools such as the Rotating-Wave approximation).

Next to $\Delta \hat b^\dagger \hat b$ in (\ref{eq:effectiveH_1}), we see that the higher orders of the effective Hamiltonian also contain diagonal terms. The collection of those additional terms proportional to $\hat b^\dagger \hat b$ in equations (\ref{eq:effectiveH_2}-\ref{eq:effectiveH_5}) is called the sub-harmonic drive-induced Stark shift
\begin{widetext}
\begin{align}
    \begin{aligned}
    \delta_3 = &-\frac{3\beta_2(2\beta_1+\beta_2)}{8\omega_\mathrm{d}}E^2\bar{\phi}^2 + \frac{5(\alpha (\beta_1+\beta_2)^2+\beta_2(2\beta_1+\beta_2)\Delta)}{32\omega_\mathrm{d}^2} E^2\bar{\phi}^2  - \frac{9(\alpha (\beta_1+\beta_2)^2(\alpha+2\Delta)+\beta_2(2\beta_1+\beta_2)\Delta^2)}{128\omega_\mathrm{d}^3}E^2\bar{\phi}^2 \\ 
    &+\frac{17(\alpha(\beta_1+\beta_2)^2(\alpha^2+3\alpha\Delta+3\Delta^2)+\beta_2(2\beta_1+\beta_2)\Delta^3)}{512\omega_\mathrm{d}^4} E^2\bar{\phi}^2 - \frac{21\beta_2^2(7\beta_1^2+10\beta_1\beta_2+4\beta_2^2)}{512\omega_\mathrm{d}^3} E^4\bar{\phi}^4 \\
    &+ \frac{(\alpha\beta_2(313\beta_1+527\beta_2)(\beta_1+\beta_2)^2+107\beta_2^2(7\beta_1^2-10\beta_1 \beta_2+4\beta_2^2)\Delta)}{2048\omega_\mathrm{d}^4} E^4\bar{\phi}^4. 
\end{aligned}
\label{app:stark_shift}
\end{align}
\end{widetext}
To avoid detuned Rabi oscillations, the effective Hamiltonian should not contain any contributions proportional to $\hat b^\dagger \hat b$. This results in the resonance condition $\Delta+\delta_3(\Delta)=0$ for sub-harmonic driving to fix $\Delta$ and by that $\omega_\mathrm{d}$.

For the present parameters, we realize that the large anharmonicity $\alpha$ suppresses leakage into the second excited state and truncate the effective model to a two-level system. Finally, time evolution is well-described by the zeroth order contribution of the Magnus expansion~\cite{magnusExponentialSolutionDifferential1954}:\footnote{Note, that in principle this formulation would also allow to treat non-trivial envelope-functions by computing higher orders in the Magnus-expansion given that the envelope varies slow enough compared to $\omega_\mathrm{d}$.}
\begin{align}
    \begin{aligned}
    \hat U(0,t_\mathrm{pulse}) &=\exp\left( -i \int_0^{t_\mathrm{pulse}} dt \hat H_{\rm eff,3}(t)\right) \\
    &= \exp ( -i t_\mathrm{pulse} \Omega_3 \hat \sigma_x),
\end{aligned}
\end{align}
with the effective Rabi rate
\begin{widetext}
    \begin{align}
    \begin{aligned}
        \Omega_3 = &\frac{1}{t_\mathrm{pulse}}\int_0^{t_\mathrm{pulse}} dt \;E(t)^3\; \left(
        \frac{\beta_1(6\alpha(\beta_1+\beta_2)^2)}{768\omega_\mathrm{d}^3}-\frac{\beta_1(6\alpha(\beta_1+\beta_2)^2(\alpha+3\Delta))}{3072 \omega_\mathrm{d}^4} 
        \right)\bar{\phi}^3 \\
        &+\frac{1}{t_\mathrm{pulse}}\int_0^{t_\mathrm{pulse}} dt \;E(t)^3\; \left(-
        \frac{\beta_1\beta_2(2\beta_1+\beta_2)}{32\omega_\mathrm{d}^2}+\frac{13\beta_1\beta_2(2\beta_1+\beta_2)\Delta}{768\omega_\mathrm{d}^3}-\frac{29\beta_1\beta_2(2\beta_1+\beta_2)\Delta^2}{3072 \omega_\mathrm{d}^4}
        \right) \bar{\phi}^3 \\
    &-\frac{1}{t_\mathrm{pulse}}\int_0^{t_\mathrm{pulse}} dt\;E(t)^5\; \frac{\beta_1\beta_2^2(29\beta_1^2+38\beta_1\beta_2+14\beta_2^2)}{1024\omega_\mathrm{d}^4} \bar{\phi}^5,
    \end{aligned}
    \end{align}
\end{widetext}
 From the structure of the commutators in (\ref{eq:order_expansion}) which contribute to the Rabi rate, it becomes obvious that only odd powers of $\phi$ can contribute; with the lowest power possible being $\phi^n$ for an $n$-photon drive. Higher powers in drive amplitude exist as well, $\Omega_n \sim c_1\bar{\phi}^{n}+c_2\bar{\phi}^{n+2}+...$, however the convergence of the power series in the experimental parameter regime results in less dominant pre-factors for larger powers in $\phi$. In total, the Rabi frequency scales polynomially with the drive power, which matches the experimental data of Fig.~\ref{fig:fig3}.

\section{Tune-Up}
\label{appendix:tune-up}
\begin{figure*}[ht]
    \centering
    \includegraphics[width=\linewidth]{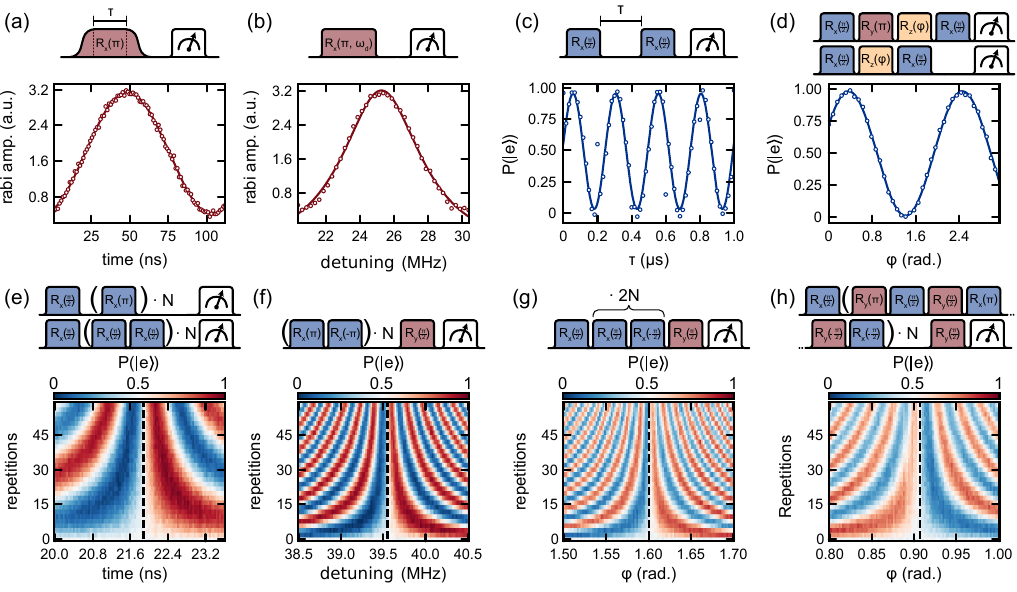}
    \caption{\textbf{Tune-up routine for single qubit gates.} The pulse is first rough-calibrated (a-d) and subsequently fine-calibrated using error amplification sequences (e-h) using different pulse sequences and subsequent measurement of the excited state population. The pulse duration and detuning from $\omega_\mathrm{eg}/n$ are calibrated with a Rabi experiment, changing the time (a) and frequency (b) of the drive pulse. The qubit frequency is calibrated using a Ramsey experiment (c). Calibration of the added pulse phase is performed separately for $\pi$ and $\pi/2$-gates (d), also using a Ramsey-type experiment with an additional virtual-Z rotation before the second $\pi/2$-gate. The error amplification sequences shown are for the pulse amplitude (e), frequency (f), $\pi/2$-pulse phase (g) and $\pi$-pulse phase (h)}
    \label{fig:tuneup}
\end{figure*}
To tune up the sub-harmonic gates, we track the qubit frequency with a local oscillator set at a fraction of the qubit frequency $\omega_\mathrm{eg}/n$, where $n$ is the photon number of the sub-harmonic used for driving. 
Translating between $X$- and $Y$-drives therefore requires changing pulse phase by $\pi/2n$, as opposed to $\pi/2$ for on-resonance driving.
As the qubit frequency is shifted upwards during driving (see Fig.~\ref{fig:fig2}), a phase shift between the qubit rotating frame and the local oscillator of the arbitrary waveform generator (AWG) arises, which is described by
\begin{equation}
    \Delta\varphi = \frac{1}{n}\int_0^{t_\mathrm{pulse}} \Delta\omega_\mathrm{eg}(t) \mathrm{d}t,
    \label{eq:phase-shift}
\end{equation}
for a pulse length $t_\mathrm{pulse}$.
Here, the shift in qubit frequency $\Delta\omega_\mathrm{eg}(t)$ includes the varying pulse amplitude over time i.e., it includes ring-up and ring-down of the pulse. 
The prefactor $1/n$ in Eq.(~\ref{eq:phase-shift}) copes with setting the AWG's local oscillator at $\omega_\mathrm{eg}/n$ to ensure a consistent phase relation between the local oscillator and the rotating frame of the qubit. 
We correct for the total phase shift~Eq. (\ref{eq:phase-shift}) acquired during a sub-harmonic gate by shifting the phase of consecutive pulses by $\Delta\varphi$, which is equivalent to a virtual-Z rotation. 

Each pulse parameter is calibrated using a two-step tune-up procedure, comprising a rough calibration pass (see Fig. \ref{fig:tuneup} a-d) and a fine calibration pass (see Fig. \ref{fig:tuneup} e-h).
We denote gates gate applied to the qubit by $R_i(\theta)$ with the rotation angle $\theta$ induced around a given axis $i\in \{x, y, z\}$.
First, the pulse length is calibrated with a Rabi experiment while sweeping the pulse time, followed by a frequency calibration with a Rabi experiment, sweeping the pulse frequency instead. 
To increase the precision of the calibration, both steps are repeated multiple times until $\omega_{\rm d}$ converges. These two steps are used to obtain the data shown in Fig. \ref{fig:fig2}\,(d) and Fig. \ref{fig:fig3}\,(b). 
Subsequently, the qubit frequency $\omega_\mathrm{eg}$ while idling is calibrated using a Ramsey pulse sequence.
To correct the additional phase added by the $\pi$- and $\pi/2$-pulses, it is calibrated separately using a Ramsey-type sequence. 
Starting with the $\pi/2$-pulse, a Ramsey sequence with zero wait time and a virtual-Z rotation with a variable phase $\varphi$ before the second pulse is used for the calibration. 
Finding the qubit in its excited state $P(|e\rangle)=1$ determines the correct value of the compensation phase $\varphi$. 
To calibrate the $\pi$-pulse, an $R_y(\pi)$-rotation is inserted into the Ramsey-sequence, making it sensitive to the added phase of the $\pi$-pulse. 
In this configuration, the sequence is insensitive to the amplitude of the $\pi$-pulse and the virtual-Z gate again compensates the added phase correctly $P(|e\rangle)=1$. 
Following the rough calibration, error amplification sequences are performed to iteratively fine-tune each pulse parameter. 
Each sequence keeps the qubit at an excited state population of 0.5 for a perfectly calibrated pulse. 
Deviating from the optimal pulse parameter causes the excited state population to oscillate, where the oscillation frequency indicates the magnitude of the error and the oscillation phase indicates the sign of the error. 
First, an $R_x(\frac{\pi}{2})$ pulse followed by $N$ sequences of $R_x(\pi)$ or $2N$ sequences of $R_x(\frac{\pi}{2})$ pulses for $\pi$ and $\pi/2$-pulses respectively \cite{sheldonCharacterizingErrorsQubit2016} amplifies the over- and under-rotation error, which is corrected by optimizing the pulse length. 
Note that the AWG has a sampling time of $t_\mathrm s=0.5$\,ns, necessitating a pulse duration of multiples of $t_\mathrm{s}$ and setting an upper bound on the precision of the gate.
This limitation is circumvented by allowing the duration of the pulse shape to change independently of the sampling rate and filling the remaining time to the next sample with a zero-amplitude pulse, greatly increasing the pulse flexibility.
Next, the frequency of the pulse is fine-tuned using a phase sensitive error amplification sequence consisting of $N$ consecutive $R_x(\pi)$ and $R_x(-\pi)$ rotation pairs~\cite{gambettaAnalyticControlMethods2011}. 
After frequency calibration, fine time calibration steps are repeated. 
Finally, the phase of the $\pi/2$ and $\pi$-pulse is calibrated with a sequence amplifying the error of the virtual-Z rotation~\cite{sheldonCharacterizingErrorsQubit2016}. 
In the sequence for the $\pi$-pulse, the qubit is first prepared on the equator of the Bloch-sphere using a $R_x(\frac{\pi}{2})$ rotation, followed by a $R(\pi)$-rotation perpendicular to the prepared state, followed by another $R_x(\frac{\pi}{2})$ rotation. 
This sequence is repeated for $R_y$-rotations, and in total repeated $N$ times.
Note that sequence (f) used for calibrating the pulse frequency is sensitive to changes in the phase-shift of the virtual-Z gate.
Therefore, the calibration routine must be executed in the exact order presented to avoid circular dependencies.
In total, starting from an un-calibrated pulse, the tune-up process is completed within 10 minutes.
\section{\label{appendix:power_scaling}Drive Amplitude Distortion}
\begin{figure}
    \centering
    \includegraphics{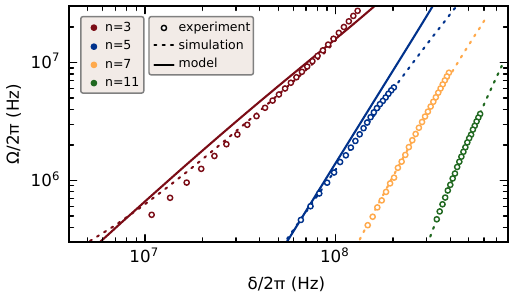}
    \caption{\textbf{Frequency shifts $\delta$ and Rabi frequencies $\Omega$ for $n^{\rm th}$ sub-harmonics}.
    Rabi frequencies as a function of drive-induced frequency shifts, as obtained from the theoretical model, numerical simulations, and experimental data.
    The expected dependence of $\Omega(\delta)$ (solid lines) from theory shows good agreement with the 3$^{\rm rd}$ sub-harmonic. Measurement errors are within the symbol size.}
    \label{fig:power_scaling}
\end{figure}
Due to the long signal path from the room temperature control electronics into the dilution cryostat, microwave signals are attenuated and distorted, making it challenging to determine the exact drive power at the qubit.
This distortion is given by the frequency-dependent transfer function of the signal line as a direct cause of impedance mismatches and the skin effect in the cables \cite{wigingtonTransientAnalysisCoaxial1957}. 
However, we find that for a sub-harmonic $n$, the drive induced frequency shift $\delta$ and therefore the on-resonance drive frequency $\omega_\mathrm{d}=\frac{1}{n}(\omega_\mathrm{eg}+\delta_n)$ is a direct measure of the drive power applied. 
In Fig. \ref{fig:power_scaling} we compare experimentally obtained Rabi frequencies vs drive-induced frequency shift for different sub-harmonics with numerical simulations and theoretical predictions. Both simulations and our model agree well with experiments, strengthening our conclusion in section \ref{coherent_control} that the transfer of the signal is the primary cause for deviations between theory and experiment. We also note that in comparing simulated vs measured frequency shifts as a function of applied power, one can extract the frequency domain transfer function of the flux line at the qubit.
\bibliography{bibliography.bib}
\end{document}